\documentclass[journal,twoside]{IEEEtran}

\ifCLASSINFOpdf
\usepackage[pdftex]{graphicx}

\else

\fi

\usepackage[cmex10]{amsmath}
\usepackage{amssymb}   
\usepackage{amsxtra}
\usepackage{amscd}
\usepackage{amsthm}
\usepackage{textcomp}
\usepackage{graphicx}

\usepackage{balance}

\usepackage{array}  

\usepackage{multirow}  

\usepackage{amsmath}
\usepackage{cite}

\usepackage{url}

\usepackage{color,soul} 
\soulregister\cite7
\soulregister\ref7
\soulregister\pageref7

\begin{document}
%

\title{{\huge Reconfigurable Intelligent Surface-Based Index Modulation: \\A New Beyond MIMO Paradigm for 6G}}


%
%
%

\author{Ertugrul~Basar,~\IEEEmembership{Senior Member,~IEEE}
            \vspace*{-0.25cm}
\thanks{Manuscript received August 5, 2019; revised December 13, 2019; accepted January 28, 2020. This work was supported in part by the Scientific and Technological Research Council of Turkey (TUBITAK) under Grant 117E869, the Turkish Academy of Sciences (TUBA) GEBIP Programme, and the Science Academy BAGEP Programme. The associate editor coordinating the review of this article and approving it for publication was Prof. J. Choi.}
\thanks{The author is with the Communications Research and Innovation Laboratory (CoreLab), Department of Electrical and Electronics Engineering, Ko\c{c} University, Sariyer 34450, Istanbul, Turkey (e-mail: ebasar@ku.edu.tr).}
\thanks{Codes available at https://corelab.ku.edu.tr/tools }
\thanks{Color versions of one or more of the figures in this article are available
online at http://ieeexplore.ieee.org.}
\thanks{Digital Object Identifier XXX}
}

\maketitle

\begin{abstract}

Transmission through reconfigurable intelligent surfaces (RISs), which control the reflection/scattering characteristics of incident waves in a deliberate manner to enhance the signal quality at the receiver, appears as a promising candidate for future wireless communication systems. In this paper, we bring the concept of RIS-assisted communications to the realm of index modulation (IM) by proposing RIS-space shift keying (RIS-SSK) and RIS-spatial modulation (RIS-SM) schemes. These two schemes are realized through not only intelligent reflection of the incoming signals to improve the reception but also utilization of the IM principle for the indices of multiple receive antennas in a clever way to improve the spectral efficiency. Maximum energy-based suboptimal (greedy) and exhaustive search-based optimal (maximum likelihood) detectors of the proposed RIS-SSK/SM schemes are formulated and a unified framework is presented for the derivation of their theoretical average bit error probability. Extensive computer simulation results are provided to assess the potential of RIS-assisted IM schemes as well as to verify our theoretical derivations. Our findings also reveal that RIS-based IM, which enables high data rates with remarkably low error rates, can become a potential candidate for future wireless communication systems in the context of beyond multiple-input multiple-output (MIMO) solutions. 
  
\end{abstract}

%
\IEEEpeerreviewmaketitle


\section{Introduction}
\IEEEPARstart{A}{s} of the first quarter of 2020, the first commercial fifth generation (5G) wireless networks have been already deployed, in part or as a whole, in certain countries while the first 5G compatible mobile devices are being introduced to the market. Although the initial 5G standard, which was completed in 2018, has brought more flexibility to the physical layer by exploiting millimeter-waves and multiple orthogonal frequency division multiplexing (OFDM) numerologies, researchers have already started to explore the potential of alternative technologies for later releases of 5G. These technologies include index modulation (IM), non-orthogonal multiple access, alternative/advanced waveforms, low-cost massive multiple-input multiple-output (MIMO) variants, terahertz communications, and new antenna technologies. At the first glance, the future 6G technologies may seem as the extension of their 5G counterparts \cite{6G}, however, new user requirements, completely new applications/use-cases, and new networking trends of 2030 and beyond may bring more challenging communication engineering problems, which necessitate radically new communication paradigms in the physical layer \cite{Saad_2019}.

Within the context of unconventional wireless communication paradigms, there has been a growing interest in controlling the reflection, scattering, and refraction characteristics of electromagnetic waves, that is, controlling the propagation, in order to increase the quality of service and/or achievable rate. IM-based  emerging schemes such as media-based modulation \cite{Khandani_conf1,MBM_TVT,Basar_2019}, spatial scattering modulation \cite{SSM}, and beam index modulation (IM) \cite{BIM}, use the variations in the signatures of received signals by exploiting reconfigurable antennas or scatterers to transmit additional information bits in rich scattering environments \cite{Basar_2017}. On the other hand, reconfigurable intelligent surfaces/walls/reflect-arrays/metasurfaces are smart devices that intentionally control the propagation environment to boost the signal quality at the receiver \cite{Akyildiz_2018,Di_Renzo_2019,Basar_Access_2019}.

As a matter of fact, the \textit{reconfigurable intelligent surface (RIS)}-based transmission concept, in which the large number of small, low-cost, and passive elements on an RIS only reflect the incident signal with an adjustable phase shift without requiring a dedicated energy source for radio frequency (RF) processing, decoding, encoding, or retransmission, is completely different from existing MIMO, beamforming, amplify-and-forward relaying, and backscatter communication paradigms. Inspired by the definition of software-defined radio, which is given as ``radio in which some or all of the physical layer functions are software defined" and considering the interaction of the intelligent surface with incoming waves in a software-defined fashion, we may also use the term of \textit{software-defined surface (SDS)} for these intelligent surfaces.  In other words, due to the fact that reflection characteristics of these intelligent surfaces/walls/arrays in the physical layer can be controlled by a software, they can be termed as SDS.

Transmission through intelligent walls is proposed in one of the early works by utilizing active frequency-selective surfaces to control the signal coverage \cite{Subrt_2012}. The promising concept of communications over smart reflect-arrays with passive reflector elements is proposed in \cite{Tan_2016} as an alternative to beamforming techniques that require large number of antennas to focus the transmitted or received signals. It has been also demonstrated that reflect-arrays can be used in an effective way to change the phase of incoming signals during smart reflection without buffering or processing them and the received signal quality can be enhanced through the adjustment of the phase shift of each reflector element on the reflect-array. The authors of \cite{Huang_2018,Huang_2018_2,Huang_2019} considered an RIS-assisted downlink transmission scenario to support multiple users and focused on the maximization of sum-rate and energy efficiency. The selection of optimum RIS phases is also investigated and low complexity algorithms are considered for the formulated non-convex optimization problems. Recently, the interesting problem of joint active and passive beamforming is investigated in \cite{Wu_2018} and \cite{Wu_2018_2}, and the user's average received power is investigated. Even more recently, the researchers focused on RIS architectures based on sparse channel sensors \cite{Taha_2019}, in which some of the existing RIS units are active, and RIS-assisted physical-layer security schemes \cite{Schober_2019_2}. 
Finally, we provided a mathematical framework in \cite{Basar_2019_LIS} for the calculation of average symbol error probability (SEP) of RIS-assisted systems. Furthermore, we proposed the promising concept of using the RIS itself as an access point (AP) by utilizing an unmodulated carrier for intelligent reflection. Interested readers are referred to \cite{Basar_Access_2019,Wu_2019,Di_Renzo_2019} and references therein for a recent overview of RIS empowered communications.

\begin{figure}[!t]
	\begin{center}
		\includegraphics[width=0.8\columnwidth]{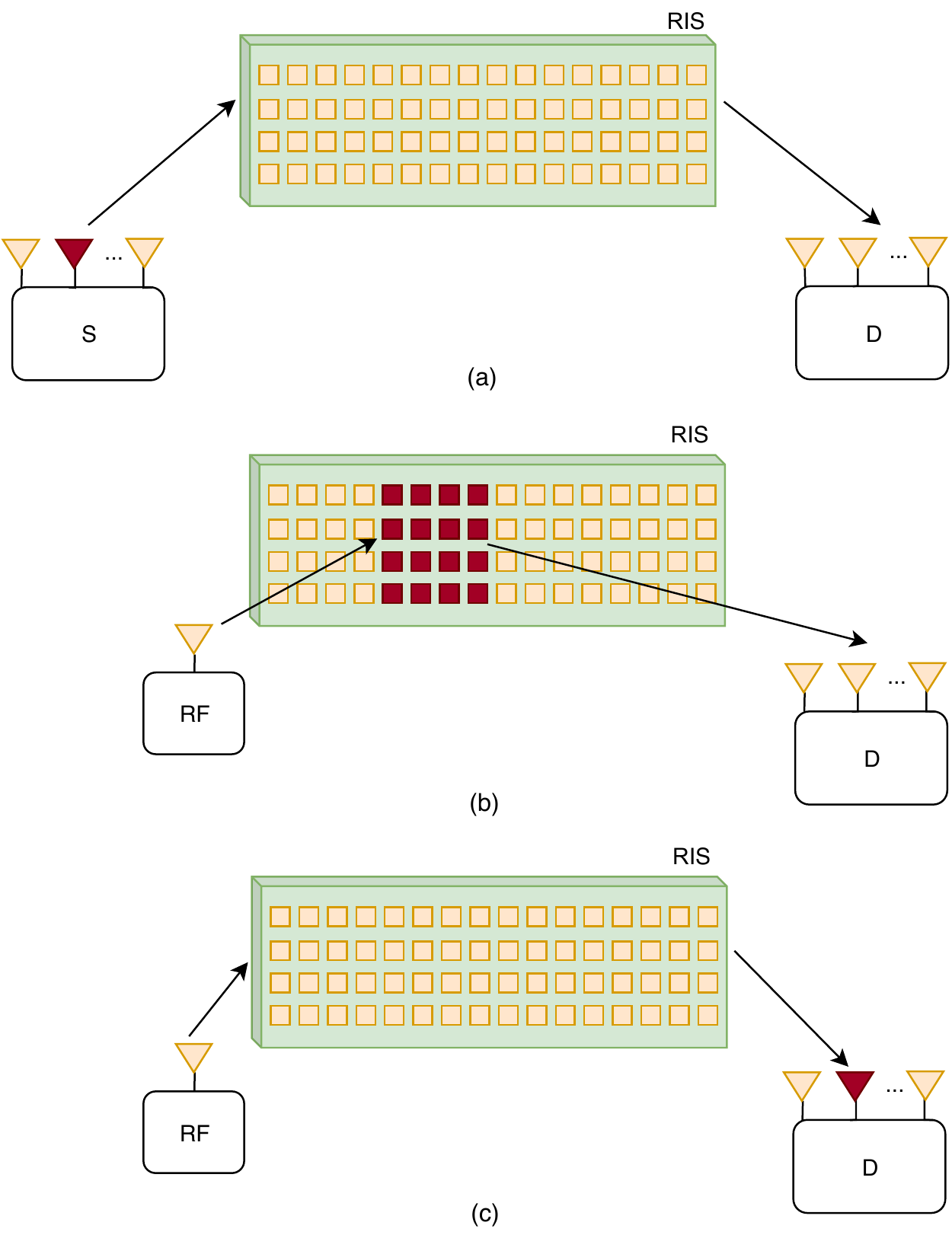}
		\vspace*{-0.3cm}\caption{Three conceptual RIS-based IM system realizations: a) IM for source (S) transmit antennas, b) IM for RIS reflector regions, c) IM for destination (D) receive antennas.}\vspace*{-0.3cm}
	\end{center}
\end{figure} 

The emerging IM concept also falls to the category of potential beyond 5G technologies and has been widely recognized by both academia and industry during the past few years \cite{IM_5G,IM_Book,Basar_2017,Mesleh_2018}. Contrary to the traditional modulation formats, the indices of the available transmit entities, such as transmit antennas for space modulation techniques \cite{Mesleh_2018} and subcarriers for OFDM with IM \cite{OFDM_IM}, are used to convey information for an IM scheme. The undeniable potential of both IM- and RIS-based communication schemes has been the main motivation of this study. With this purpose, we investigated three conceptual RIS-based IM system realizations in Figs. 1(a)-(c), in which we consider IM for source (S) transmit antennas, RIS regions, and destination (D) receive antennas, respectively. Since the first concept requires the knowledge of activated transmit antenna indices of S at RIS for optimum reflection, i.e., requires an extra signaling link between S and the RIS, and the second concept reduces the effective gain of the RIS by activating a part of the available reflectors, i.e., reduces the effective received signal power, we decided to focus on the third approach in this preliminary work.

In this paper, we propose the visionary concept of \textit{RIS-based IM} as a potential beyond MIMO solution by amalgamating the techniques of transmission over RISs and IM for receive antenna indices to achieve high reliability along with high spectral efficiency. Different from emerging massive MIMO systems with fully digital or hybrid beamforming, our design neither requires multiple RF chains nor analog phase shifters at transmitter/receiver sides and exploits the inherent randomness of the propagation environment. First, considering the RIS as an AP \cite{Basar_2019_LIS,Basar_Access_2019}, we propose RIS-space shift keying (RIS-SSK) and RIS-spatial modulation (RIS-SM) schemes by exploiting the RIS not only to boost the signal quality in hostile fading channels but also to realize IM by the selection of a particular receive antenna index according to the information bits. Second, we formulate the greedy and maximum likelihood (ML) detectors of both schemes and investigate their complexity. Third, we present a unified framework for the calculation of the theoretical error performance of the proposed schemes and provide useful insights. Finally, extensive computer simulations are given to assess the potential of the RIS-SSK and RIS-SM schemes.

The rest of the paper is organized as follows. In Section II, we introduce the system model of RIS-based SSK/SM schemes and formulate their detectors. Sections III and IV respectively focuses on the error performance of greedy and ML detectors. Computer simulation results and comparisons are given in Section V. Finally, conclusions are given in Section VI\footnote{\textit{Notation}: Bold, lowercase and capital letters are used for column vectors and matrices, respectively. $(\cdot)^*$ $(\cdot)^\mathrm{T}$, and $(\cdot)^\mathrm{H}$ denote complex conjugation, transposition, and Hermitian transposition, respectively. The real and imaginary parts of a complex variable $X$ are denoted by $X_{\Re}$ (or $\Re \left\lbrace X \right\rbrace $) and $X_{\Im}$ (or $\Im \left\lbrace X \right\rbrace $), respectively. $\mathrm{det}(\cdot)$ and $(\cdot)^{-1}$ stand for the determinant and the inverse of a matrix, and $\mathrm{diag}(\cdot)$ returns a diagonal matrix from a vector. The $n\times n$ identity matrix is denoted by $\mathbf{I}_n$. $X\sim \mathcal{N}(\mu,\sigma^2)$ stands for the real Gaussian distribution of $X$ with mean $\mathrm{E}[X]=\mu$ and variance $\mathrm{Var}[X]=\sigma^2$, while $\mathcal{CN}(0,\sigma^2)$ represents circularly symmetric complex Gaussian distribution with variance $\sigma^2$. $P(\cdot)$ stands for the probability of an event. $f_X(x)$, $F_X(x)$, $\Psi_X(w)$, and $M_X(s)$ respectively stand for the probability density function (PDF), cumulative distribution function (CDF), characteristic function (CF), and moment generating function (MGF) of a random variable (RV) $X$. $Q(\cdot)$ is the Gaussian $Q$-function and $j=\sqrt{-1}$.
}.

\begin{figure}[!t]
	\begin{center}

		\includegraphics[width=1.0\columnwidth]{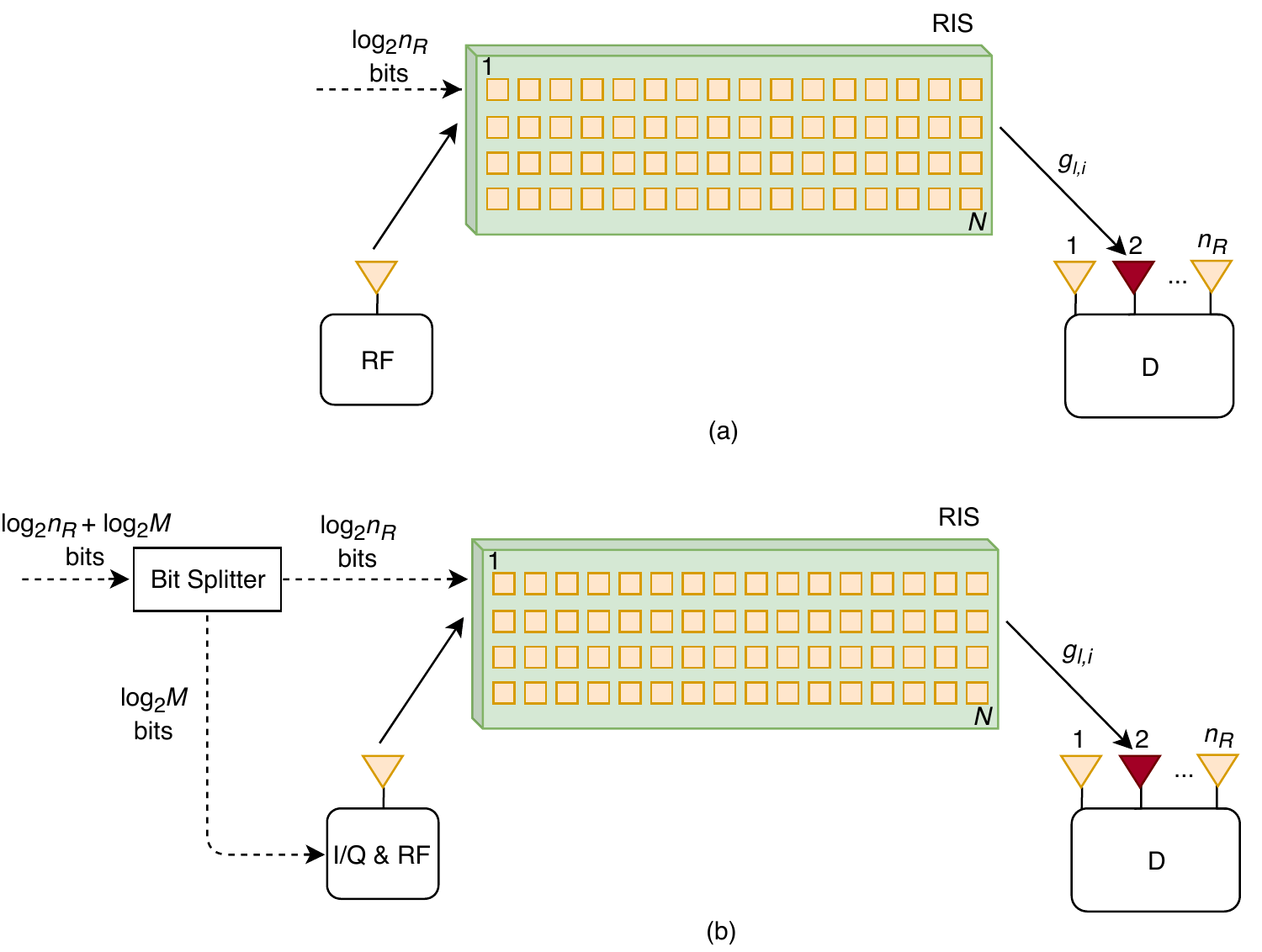}
		\vspace*{-0.4cm}\caption{RIS-based IM schemes: a) RIS-SSK, b) RIS-SM.}\vspace*{-0.3cm}
	\end{center}
\end{figure}

\vspace*{-0.3cm}
\section{System Model}
In this section, we present the system models of the proposed RIS-based SSK and SM schemes and investigate the problem of optimal RIS phase selection to achieve the best error performance. We build the proposed schemes on the concept of RIS-AP introduced in \cite{Basar_2019_LIS,Basar_Access_2019}, where the RIS, which is a part of the transmitter, reflects the signals generated by a near RF source in a deliberate manner to convey information bits. We assume that the RIS is consisting of $N$ passive and low-cost reflector elements (reconfigurable meta-surfaces), while D, which lies in the far-field of the RIS and does not receive transmission from the RF source, is equipped with $n_R$ receive antennas. The wireless fading channel between the $l$th receive antenna of D and $i$th reflector element is characterized by $g_{l,i}=\beta_{l,i} e^{-j \psi_{l,i}}$ for $l=1,2,\ldots,n_R$ and $i=1,2,\ldots,N$, and follows $\mathcal{CN}(0,1)$ distribution under the assumption of flat Rayleigh fading channels \cite{Huang_2019}. We also assume that all wireless channels are uncorrelated and perfect channel state information (CSI) is available at D if it is required by the utilized detector. It is worth noting that more practical channel modeling and phase shift design \cite{Abeywickrama_2019} along with effective channel estimation techniques \cite{He_2019} are open research topics with the context of RISs and the considered model may shed light on performance limits of future schemes. Particularly, some level of spatial correlation is likely to
be expected among reflector elements and this may affect the ultimate performance limits. The phase induced by the $i$th reflector is shown by $\phi_i$ for $i=1,2,\ldots,N$. For intelligent reflection, the RIS has the knowledge of channel phases $\psi_{l,i}$ for all $l$ and $i$, which is a common assumption in the literature to model the optimum system behavior \cite{Huang_2018,Huang_2018_2,Huang_2019,Wu_2018,Wu_2018_2}.

\subsection{RIS-Assisted Space Shift Keying}
In this scenario, the unmodulated carrier signal generated by the RF source, is reflected to D with the aim of maximizing the instantaneous received SNR at a specific receive antenna, which is selected according to the incoming $\log_2 n_R$ information bits, as shown in Fig. 2(a). In other words, the RIS phase terms are adjusted in such a way that the SNR at the target receive antenna is maximized, while the task of D is to detect the index of its receive antenna with the maximized instantaneous received SNR.

For this system, the transmitted signal is reflected by an RIS that is made
of $ N $ reflecting elements. Ignoring the fading effects between the RF source and the RIS, the received baseband signal at the $l$th receive antenna of D is given as
\begin{equation}\label{1}
r_l=\sqrt{E_s} \left[ \sum_{i=1}^{N} g_{l,i} e^{j\phi_i} \right] + n_l, \quad l\in \left\lbrace 1,\ldots,n_R \right\rbrace
\end{equation}
where $E_s$ is the transmitted signal energy of the unmodulated carrier and $n_l$ is the additive white Gaussian noise (AWGN) sample at the $l$th receiver, which follows $\mathcal{CN}(0,N_0)$ distribution. Here, the reflector phases $\left\lbrace \phi_i\right\rbrace _{i=1}^N$ are adjusted according to the information bits to maximize the received SNR at a specific receive antenna. For this purpose, the incoming $\log_2 n_R$ bits specify the index $m$ of a receive antenna and the RIS adjusts its phases according to this selected receive antenna as $\phi_i=\psi_{m,i}$ for $i=1,2,\ldots,N$. More specifically, the received instantaneous SNR at the $l$th receive antenna is expressed as
\begin{equation}\label{2}
\gamma_l = \frac{\big|  \sum_{i=1}^{N} \beta_{l,i} e^{j(\phi_i-\psi_{l,i})} \big|^2 E_s }{N_0}.
\end{equation} 
Considering 

\small
\vspace*{-0.4cm}
\begin{equation}\label{3}
\left| \sum_{i=1}^{N} z_i e^{j \xi_i}\right| ^2 = \sum_{i=1}^{N} z_i^2 + 2 \sum_{i=1}^{N} \sum _{k=i+1}^N  z_i z_k \cos(\xi_i - \xi_k) 
\end{equation}
\normalsize
which can be maximized by ensuring $\xi_i=\xi$ for all $i$, to maximize the instantaneous SNR at the $m$th receive antenna, we select  $\phi_i=\psi_{m,i}$. This results in the following maximum SNR value at the selected receive antenna:
\begin{equation}\label{4}
\gamma_m = \frac{\big|  \sum_{i=1}^{N} \beta_{m,i} \big|^2 E_s }{N_0}.
\end{equation}
Later, we will also show that this selection of RIS phases is optimal in terms of error performance. We introduce two different detectors for the RIS-SSK scheme, which are commonly used for existing IM-based systems.

\textit{i) Greedy Detector}: This simple yet effective detector eliminates the need for channel estimation at D, that is, performs non-coherent detection, and simply detects the selected receive antenna as the one with the highest instantaneous energy:
\begin{equation}\label{5}
\hat{m} = \arg \max_{m} \left| r_m \right|^2. 
\end{equation} 
As seen from \eqref{5}, this detector does not require CSI, which can be prohibitive at D for increasing $N$ and $n_R$.

In order to shed light on the derivation of the optimum RIS phases, let us consider the selection of the $m$th receive antenna of D at the RIS and its erroneous detection as $\hat{m}$. For this case, the corresponding pairwise error probability (PEP) can be easily expressed from \eqref{5} as

\small
\vspace*{-0.4cm}
\begin{align}\label{6}
P(m \rightarrow \hat{m}) &= P ( \left|r_m \right|^2 < \left|r_{\hat{m}} \right|^2 ) \nonumber \\
&\hspace*{-1.8cm}= \!P \!\left( \left|\! \sqrt{E_s}  \sum_{i=1}^{N}\! g_{m,i} e^{j\phi_i} \! + \!n_m \right|^2 \!\! \! < \left|\! \sqrt{E_s}  \sum_{i=1}^{N} \! g_{\hat{m},i} e^{j\phi_i} \! +\! n_{\hat{m}} \right|^2   \right) 
\end{align}
\normalsize
where $\left\lbrace \phi_i\right\rbrace _{i=1}^N$ is determined according to the $m$th receive antenna (with a predefined method). As seen from \eqref{6}, a logical selection of $\phi_i$'s should minimize this PEP for the selected receive antenna $m$. The above phase optimization problem can be reformulated as follows by ignoring the noise terms:

\small
\vspace*{-0.4cm}
\begin{equation}\label{7}
\min_{\left\lbrace \phi_i\right\rbrace _{i=1}^N}  \!  \! P\left(\left|  \sum_{i=1}^{N} \! \beta_{m,i} e^{j (\phi_i-\psi_{m,i})} \right|^2  \! \! \!< \left|  \sum_{i=1}^{N} \! \beta_{\hat{m},i} e^{j (\phi_i-\psi_{\hat{m},i})} \right|^2 \right)\!.
\end{equation} 
\normalsize
As seen from \eqref{7}, even for a specific pair of $m$ and $\hat{m}$, this optimization is not a straightforward task. Consequently, we aim to maximize the first term by letting $\phi_i=\psi_{m,i}$ for all $i$ to minimize this probability. We will also show later that this selection maximizes the mean of the first term in \eqref{7}, while providing a zero-mean for the second one.

\textit{ii) ML Detector}: The ML detector of the RIS-SSK scheme considers the received signals at all receive antennas of D and performs the detection as follows:
\begin{equation}\label{8}
\hat{m}=\arg \min_m \sum_{l=1}^{n_R}\left| r_l-\sqrt{E_s} \left[ \sum_{i=1}^{N} g_{l,i} e^{j \psi_{m,i}} \right] \right| ^2.
\end{equation}   
In terms of computational complexity, comparing \eqref{5} and \eqref{8}, we observe that the ML detector requires not only CSI but also $\sim\!\! Nn_R^2$ real multiplications (RMs), while the greedy detector requires only $\!n_R$ squared complex modulus operations ($\sim n_R$ RMs). As we will show in later sections, the price paid for this increased complexity can be compensated with the improved error performance.

\subsection{RIS-Assisted Spatial Modulation}
For the RIS-SM scheme, ordinary $M$-ary modulation formats are also considered at the RF source to further improve the spectral efficiency. As shown in Fig. 2(b), the incoming $\log_2 n_R + \log_2 M$ information bits are partitioned into two groups. While the first group of $\log_2 n_R $ bits adjusts the RIS phases according to the selected receive antenna with index $m$ as done for the RIS-SSK scheme, i.e., $\phi_i=\psi_{m,i}$ for $i=1,2,\ldots,N$, the second group of $\log_2 M $ bits is passed to the RF source for the generation of an amplitude/phase modulated signal through an RF chain. Consequently, the received signal at the $l$th receive antenna of D is expressed as
\begin{equation}\label{9}
r_l= \left[ \sum_{i=1}^{N} g_{l,i} e^{j\phi_i} \right]x + n_l, \quad l\in \left\lbrace 1,\ldots,n_R \right\rbrace
\end{equation} 
where $x$ is the data symbol selected from $M$-QAM/PSK constellations, $\mathrm{E}[\left|x \right|^2 ]=E_s$, and $n_l \sim \mathcal{CN}(0,N_0)$ is the noise term. In a similar way, we propose the greedy and ML detectors of RIS-SM in the sequel.

\textit{i) Greedy Detector}: This detector simplifies the receiver design by detecting the selected receive antenna index and the transmitted symbol in a sequential fashion. For this purpose, the selected receive antenna is determined similar to the RIS-SSK scheme: $\hat{m} = \arg \max_{m} \left| r_m \right|^2$. After the detection of the selected receive antenna index as $\hat{m}$, this detector demodulates the transmitted data symbol as
\begin{equation}\label{10}
\hat{x} = \arg \min_x \left|r_{\hat{m}} - \left(\sum_{i=1}^{N} \beta_{\hat{m},i} \right) x   \right|^2  .
\end{equation}  
Here, compared to RIS-SSK, an additional (but minor) complexity comes during the search for the constellation point $x$, however, due to disjoint detection of $m$ and $x$, the overall complexity still linearly increases with $n_R$ and $M$. Furthermore, this detector requires only channel amplitudes for detection. It is worth noting that for constant-envelope constellations such as $M$-PSK, the transmitted symbol can be detected even without channel amplitudes since $\sum_{i=1}^{N} \beta_{\hat{m},i} $ is a real variable:
\begin{equation}\label{11}
\hat{x} = \arg \min_x \left|r_{\hat{m}} - x   \right|^2.
\end{equation} 

\textit{ii) ML Detector}: This detector performs a joint search for the selected receive antenna index $m$ and the transmitted data symbol $x$ by considering all received signals as 
\begin{equation}\label{12}
(\hat{m},\hat{x})=\arg \min_{(m,x)} \sum_{l=1}^{n_R}\left| r_l- \left[ \sum_{i=1}^{N} g_{l,i} e^{j \psi_{m,i}} \right]x \right| ^2.
\end{equation}
As seen from \eqref{12}, the ML detector of the RIS-SM scheme requires the full CSI along with $ \sim \!(N+M) n_R^2$ RMs to make a joint decision on $(m,x)$, while its greedy detector requires only $\sim (n_R+M)$ RMs. 

Finally, it is worth noting that for large $N$ and $n_R$, the greedy detector would be a more practical choice since it does not require dedicated channel estimation for D.

\section{Greedy Detection: Performance Analysis}
In this section, we investigate the theoretical bit error probability (BEP) of the proposed RIS-SSK and RIS-SM schemes in the presence of greedy detection. We also provide useful insights regarding the asymptotic behavior of the proposed schemes with this type of detection.

\subsection{Performance of RIS-SSK}
Based on the detection rule given in \eqref{5}, the corresponding PEP is given in \eqref{6} for the erroneous detection of the selected receive antenna index $ m $ as $\hat{m}$. Considering  $\phi_i=\psi_{m,i}$ for $i=1,2,\ldots,N$, \eqref{6} simplifies to
\begin{align}\label{13}
P(m \rightarrow \hat{m})= P \!\left( \left| \sqrt{E_s}  B  + n_m \right|^2 \!\! < \! \left|\! \sqrt{E_s}  \hat{B} + n_{\hat{m}} \right|^2   \right) 
\end{align}
where $B=\sum_{i=1}^{N}\! \beta_{m,i}$ and  $\hat{B}=\sum_{i=1}^{N}  \beta_{\hat{m},i} e^{j(\psi_{m,i}- \psi_{\hat{m},i})}$. Here, we resort to the central limit theorem (CLT) under the assumption of $N \gg 1$ for the calculation of this PEP. Under the CLT, $B$ and $\hat{B}$ follow Gaussian distribution regardless of the distributions of their components. Specifically, since $\beta_{m,i}$ is a Rayleigh distributed RV with $\mathrm{E}[\beta_{m,i}]=\sqrt{\pi}/2$ and $\mathrm{Var}[\beta_{m,i}]=(4-\pi)/4$, we have $B\sim \mathcal{N}(N\sqrt{\pi}/2,N(4-\pi)/4)$. Since $\psi_{m,i}$ and $\psi_{\hat{m},i}$ are independent and uniformly distributed in $(0,2\pi)$, the distribution of $ \bar{\psi_i}=\psi_{m,i}- \psi_{\hat{m},i}$ is obtained as
\begin{equation}\label{14}
f_{\bar{\psi_i}}(x)=\begin{cases}
\frac{1}{2\pi} \left(1+\frac{x}{2 \pi} \right), \quad -2\pi<x<0 \\
\frac{1}{2\pi} \left(1-\frac{x}{2 \pi} \right), \quad\quad\,\, 0<x<2\pi. 
\end{cases}
\end{equation}
Then defining $\hat{B}_i=\beta_{\hat{m},i} e^{ j\bar{\psi_i}}$, we have $\mathrm{E}[\hat{B}_i]=0$ and $\mathrm{Var}[(\hat{B}_i)_{\Re}]=\mathrm{Var}[(\hat{B}_i)_{\Im}]=0.5$, due to the symmetry of cosine and sine functions. As a result, we obtain $\hat{B} \sim \mathcal{CN}(0,N)$ with independent and identically distributed (iid) real and imaginary parts. In light of this information, we obtain
\begin{align}\label{15}
(\sqrt{E_s}  B  + n_m)_{\Re} & \sim \mathcal{N} \big( \tfrac{N\sqrt{\pi E_s}}{2},\tfrac{N(4-\pi)E_s}{4}+\tfrac{N_0}{2}\big)  \nonumber \\
(\sqrt{E_s}  B  + n_m)_{\Im} & \sim \mathcal{N} \left( 0,\tfrac{N_0}{2}\right)  \nonumber \\
(\sqrt{E_s}  \hat{B} + n_{\hat{m}})_{\Re} & \sim \mathcal{N}\left(0,\tfrac{NE_s+N_0}{2} \right)   \nonumber \\
(\sqrt{E_s}  \hat{B} + n_{\hat{m}})_{\Im} & \sim \mathcal{N}\left(0,\tfrac{NE_s+N_0}{2} \right).  
\end{align}
Considering \eqref{15}, we may re-express \eqref{13} as
\begin{equation}\label{16}
P(m \rightarrow \hat{m})= P(Y<0)=P(Y_1 + Y_2 - Y_3<0)
\end{equation}
where $Y_1=(\sqrt{E_s}  B  + n_m)_{\Re}^2$ is non-central chi-square $(\chi^2)$ RV with one degree of freedom, $Y_2=(\sqrt{E_s}  B  + n_m)_{\Im}^2$ is a central $\chi^2$ RV with one degree of freedom, and $Y_3= \big|\! \sqrt{E_s}  \hat{B} + n_{\hat{m}} \big|^2$ is a central $\chi^2$ RV with two degrees of freedom. Due to the complexity of the distribution of $Y_1+Y_2$ \cite[Eq. (5.45)]{Simon2}, we use the Gil-Pelaez's inversion formula \cite[Eq. (4.4.1)]{Mathai_1992}
\begin{equation}\label{17}
F_Y(y)=\frac{1}{2}-\int_{0}^{\infty} \frac{\Im\left\lbrace   e^{-jw y} \Psi_Y(w) \right\rbrace    }{w \pi} dw
\end{equation} 
where $F_Y(y)=P(Y\le y)$ is the CDF and $\Psi_Y(w) =\mathrm{E}[e^{jwY}]$ is the CF of $Y$. Since the CF of the sum of independent RVs (thanks to zero-mean RVs in \eqref{15}, we satisfy this condition) is the multiplication of their individual CFs, we obtain $\Psi_Y(w)=\Psi_{Y_1}(w) \Psi_{Y_2}(w) \Psi_{(-Y_3)}(w)$. Consider the generic CF of a non-central $\chi^2$ RV $ X $ with $n$ degrees of freedom, which is given as
\begin{equation}\label{18}
\Psi_X(w)=\left(\frac{1}{1-2jw \sigma^2} \right)^{n/2} \exp\left(\frac{jw\mu^2}{1-2jw \sigma^2} \right) 
\end{equation}
where $X=\sum_{k=1}^{n} X_k^2$ and $X_k \sim \mathcal{N}(\mu_k,\sigma^2)$ and $\mu^2 = \sum_{k=1}^{n} \mu_k^2 $. It is worth noting that for a central $\chi^2$ RV, we have $\mu=0$ in \eqref{18}. Substituting the values of \eqref{15} in \eqref{18} for $Y_1,Y_2$, and $Y_3$, considering $\Psi_{(-Y_3)}(w)=\Psi_{(Y_3)}(-w)$, and evaluating the integral in \eqref{17} numerically\footnote{For numerical integration, the infinity in the upper limit of the integral in \eqref{17} is replaced by $10^3$ to avoid numerical calculation errors.} for $y=0$,   we obtain the corresponding exact PEP, i.e., $P(m \rightarrow \hat{m})=F_Y(0)$. 

To gain further insights, considering the fact that $\mathrm{E}[Y_1] \gg \mathrm{E}[Y_2] $ for large $N$, the PEP in \eqref{16} can be upper bounded by
\begin{equation}\label{19}
P(m \rightarrow \hat{m})<P(Y_1 - Y_3<0).
\end{equation}
Defining $\tilde{Y}=Y_1-Y_3$ and considering the PDF of the difference of a non-central and a central $\chi^2$ RV \cite[Eq. (4.35)]{Simon2}, we obtain
\begin{equation}\label{20}
P(m \rightarrow \hat{m})<\!\int_{-\infty}^{0} \!\! \! \! f_{\tilde{Y}}(\tilde{y}) d\tilde{y}=\exp\left(\frac{-m_1^2}{2(\sigma_1^2+\sigma_2^2)} \right) \sqrt{\frac{\sigma_2^2}{\sigma_1^2+\sigma_2^2}} 
\end{equation}
where $m_1=\frac{N\sqrt{\pi E_s}}{2}$, $\sigma_1^2=\frac{N(4-\pi)E_s}{4}+\frac{N_0}{2}$ and $\sigma_2^2=\frac{NE_s+N_0}{2}$. Simple manipulations give
\begin{equation}\label{21}
P(m \rightarrow \hat{m})<  \left( \frac{1+\frac{NE_s}{N_0}}{2+\frac{NE_s(6-\pi)}{2N_0}} \right) ^{\!\frac{1}{2}} \exp\left( \frac{\frac{-N^2 \pi E_s}{N_0}}{8+\frac{2NE_s(6-\pi)}{N_0}} \right).
\end{equation}
According to \eqref{21}, for the SNR range of interest $\frac{NE_s}{N_0} \ll 10$ (although the SNR ($E_s/N_0$) is relatively small, considerably low BEP values are possible for RIS-based systems in this region with increasing $N$ \cite{Basar_2019_LIS}), we obtain 
\begin{equation}\label{22}
P(m \rightarrow \hat{m}) \propto \exp \left( \frac{-N^2 \pi E_s}{8 N_0} \right) 
\end{equation}
which suggests a superior index detection probability for the RIS-SSK scheme with increasing $N$.  

It is important to note that for binary signaling, i.e., $n_R=2$, the above PEP yields the exact BEP. However, for the general case of $n_R>2$, we use the following union bound:
\begin{align}\label{23}
P_b &\!\le\! \frac{1}{\log_2 \! n_R}\! \sum_{\hat{m}}\! P(m \rightarrow \hat{m}) e(m,\hat{m}) \!=\! \frac{n_R}{2}  P(m \rightarrow \hat{m})
\end{align}
where $e(m,\hat{m})$ is the Hamming distance between the binary representations of $m$ and $\hat{m}$. Here, we considered the fact that the resulting PEP is independent of $m$ and $\hat{m}$, and identical for all pairs (uniform error probability) and $\sum_{\hat{m}} e(m,\hat{m})=(n_R/2) \log_2 n_R$ for all $m$ due to bit symmetry. For simplicity, we adopt natural mapping for receive antenna indices.

\textit{Remark 1}: We observe from the exact and upper-bounded PEP expressions of \eqref{16} and \eqref{22} that the resulting PEP is independent of $n_R$ for greedy detection of RIS-SSK. As seen from \eqref{23}, doubling $n_R$ doubles $P_b$ in high SNR.

\subsection{Performance of RIS-SM}
To derive the theoretical BEP of the RIS-SM scheme with greedy detection, we consider the following approximation:
\begin{equation}\label{24}
P_b \approx \frac{P_c(m)P_s}{\log_2 (Mn_R)} + 0.5 P_e(m)
\end{equation}
where $P_c(m)$ is the average correct detection probability of the selected receive antenna with index $m$ for RIS-SM, which is the same for all $m$, $P_s$ is the average symbol error probability (SEP) conditioned on correct index detection, and $P_e(m)=1-P_c(m)$ is the erroneous index detection probability. Here, we followed a conservative approach by assuming that approximately $50\%$ of the transmitted bits are erroneously detected if the receive antenna index is erroneously estimated, which is a valid assumption for IM-based systems due to error propagation. Resorting to the union bound on error probability with uniform PEP values, we obtain
\begin{equation}\label{25}
P_e(m) \le \sum_{\hat{m}=1, \hat{m}\neq m}^{n_R} \!\! \bar{P}(m \rightarrow \hat{m})=(n_R-1) \bar{P}(m \rightarrow \hat{m})
\end{equation} 
where $\bar{P}(m \rightarrow \hat{m})$ is the PEP associated with index detection averaged over all data symbols for RIS-SM. From \eqref{25}, $P_c(m)$ is obtained as $P_c(m) \ge 1-(n_R-1) \bar{P}(m \rightarrow \hat{m}) $. 

Assuming that $x$ is transmitted, the PEP for  erroneous detection of the selected receive antenna index $m$ as $\hat{m}$ is given as follows considering \eqref{5}, which is also valid for RIS-SM:
\begin{align}\label{26}
P(m \rightarrow \hat{m} \!\left.\right|\! x)= P \!\left( \big|  Bx  + n_m \big|^2 < \big|   \hat{B}x + n_{\hat{m}} \big|^2   \right). 
\end{align}
Here, $B$ and $\hat{B}$ are as defined in \eqref{13}. In what follows, we calculate \eqref{26} for different constellations.

\textit{i) BPSK}: For this case, we have $x \in \left\lbrace \pm \sqrt{E_s}\right\rbrace$ and we obtain the same PEP derived from \eqref{17} or \eqref{21} for RIS-SSK, which is independent of $x$.

\textit{ii) $M$-QAM}: For this case, we have $x=x_{\Re}+jx_{\Im}$ with $\mathrm{E}[\left| x\right|^2   ]=E_s$, and we may express the corresponding PEP as
\begin{equation}\label{27}
P(m \rightarrow \hat{m}\!\left.\right|\! x)=P(B_1^2+B_2^2-B_3^2-B_4^2<0)=P(D<0)
\end{equation}   
where $B_1=(Bx+n_m)_{\Re}$, $B_2=(Bx+n_m)_{\Im}$, $B_3=(  \hat{B}x + n_{\hat{m}} )_{\Re}$, and $B_4=(  \hat{B}x + n_{\hat{m}} )_{\Im}$, and all follow Gaussian distribution. Unfortunately, $B_1$ and $B_2$ are correlated through $B$ due to non-zero values of $x_{\Re}$ and $x_{\Im}$, and we have to express $D$ in the quadratic form of Gaussian RVs to derive its statistics. Considering $D=\mathbf{x}^\mathrm{T}\mathbf{A}\mathbf{x}$ for $\mathbf{x}=\begin{bmatrix}
B_1 & B_2 & B_3 & B_4
\end{bmatrix}^\mathrm{T}$ and $\mathbf{A}=\mathrm{diag}(\begin{bmatrix}
1 & 1 & -1 & -1
\end{bmatrix})$, the mean vector and covariance matrix of $\mathbf{x}$ are respectively calculated as
\begin{equation}\label{28}
\mathbf{m}=\begin{bmatrix}
\frac{N\sqrt{\pi}x_{\Re}}{2} & \frac{N\sqrt{\pi}x_{\Im}}{2} & 0 & 0
\end{bmatrix}^\mathrm{T}
\end{equation}
\begin{equation}\label{29}
\mathbf{C}=\!\begin{bmatrix}
\frac{N(4-\pi)x_{\Re}^2}{4} \!+\! \frac{N_0}{2} & \sigma_{1,2} & 0 & 0 \\
\sigma_{1,2} & \frac{N(4-\pi)x_{\Im}^2}{4}\!+\! \frac{N_0}{2} & 0 & 0 \\
0 & 0 & \frac{NE_s+N_0}{2} & 0 \\
0 & 0 & 0 & \frac{NE_s+N_0}{2}
\end{bmatrix}
\end{equation}
where $\sigma_{1,2}$ is the covariance of $B_1$ and $B_2$, given as
\begin{equation}\label{30}
\sigma_{1,2}=\mathrm{E}[B_1B_2]-\mathrm{E}[B_1]\mathrm{E}[B_2]=
 \tfrac{N(4-\pi) x_{\Re} x_{\Im} }{4} . 
\end{equation}
Then the MGF of $ D $ can be calculated from \cite[Eq. (3.2a.1)]{Mathai_1992} as
\begin{align}\label{31}
M_D(s)&=(\det( \mathbf{I}-2s\mathbf{AC})) ^{-\frac{1}{2}} \nonumber\\
&\hspace{-0.2cm}\times \exp\left(-\frac{1}{2} \mathbf{m}^\mathrm{T} \left[\mathbf{I}-(\mathbf{I}-2s\mathbf{AC})^{-1}  \right] \mathbf{C}^{-1} \mathbf{m}  \right). 
\end{align}
The CF of $D$ $(\Psi_D(w))$ can be obtained by replacing $s$ by $jw$ in \eqref{31}. Finally, Gil-Pelaez's inversion formula \eqref{17} can be used to calculate the PEP.

Due to the symmetry of the real and imaginary parts of $x$ for QPSK ($ 4 $-QAM), i.e., $x_{\Re}^2=x_{\Im}^2=E_s/2$, it can be proved that the above PEP is independent of $x$. However, for $M$-QAM with $M>4$, $P(m \rightarrow \hat{m}\!\left.\right|\! x)$ becomes dependent on $x$ and the average PEP for index detection can be obtained via
\begin{equation}\label{32}
\bar{P}(m \rightarrow \hat{m}) = \frac{1}{M} \sum_x P(m \rightarrow \hat{m} \!\left.\right|\! x).
\end{equation}
Substitution of \eqref{32} in \eqref{25} yields an upper bound on the average erroneous index detection probability $P_e(m)$.  

For the calculation of the average SEP ($P_s$) under the condition of correct index detection, we may consider the following equivalent signal model from \eqref{9}:
\begin{equation}\label{33}
r_m= \left[ \sum_{i=1}^{N} \beta_{m,i}  \right]x + n_m = Bx+n_m,
\end{equation}  
where the MGF of the instantaneous received SNR $\gamma=E_s B^2 /N_0$ is derived in \cite[Eq. (19)]{Basar_2019_LIS},  as
\begin{equation}\label{34}
M_{\gamma}(s)= \left( \dfrac{1}{1-\frac{sN(4-\pi)E_s}{2N_0}}\right) ^{ \! \frac{1}{2}}  \! \!\exp\left( \dfrac{ \frac{sN^2 \pi E_s}{4 N_0}}{1-\frac{sN(4-\pi)E_s}{2N_0}} \right). 
\end{equation}
Using this MGF, $P_s$ can be easily calculated for BPSK and square $M$-QAM constellations \cite{Simon}, respectively, as 
\begin{equation}\label{35}
P_s=\frac{1}{\pi} \int_{0}^{\pi/2} M_{\gamma} \left( \frac{-1}{\sin^2 \! \eta}\right) d\eta,
\end{equation}
\vspace*{-0.3cm}
\begin{align}\label{36}
P_s = \,\, & \, \frac{4}{\pi} \left(1-\frac{1}{\sqrt{M}} \right) \int_{0}^{\pi/2} M_{\gamma} \left(  \frac{-3}{2(M-1) \sin^2 \! \eta}\right) d\eta  \nonumber \\
&\hspace*{-0.7cm}- \frac{4}{\pi} \left(1-\frac{1}{\sqrt{M}} \right)^2 \int_{0}^{\pi/4} M_{\gamma} \left(  \frac{-3}{2(M-1) \sin^2 \! \eta}\right) d \eta.  
\end{align}
Finally, substituting \eqref{25} and \eqref{35} (or \eqref{36}) in \eqref{24} provides the desired $P_b$.

\textit{Remark 2}: Our observations indicate that for high SNR, which is relative for RIS-based schemes, $P_c(m) P_s \ll P_e(m)$, and $P_b$ is dominated by $P_e(m)$, i.e., $P_b \propto P_e(m)$, which deteriorates with increasing $n_R$ as well.

\section{Maximum Likelihood Detection: Performance Analysis}
In this section, we extend our theoretical analyses to the ML detection of RIS-SSK and RIS-SM schemes. To provide a more concise and intuitive presentation, we first deal with the RIS-SM scheme and then extend our theoretical derivations to RIS-SSK by simply assuming $x=\hat{x}=\sqrt{E_s}$. 

\subsection{Performance of RIS-SM}
To derive the theoretical BEP of the RIS-SM scheme for ML detection, we consider the underlying PEP for joint detection of the selected receive antenna index $m$ and the transmitted data symbol $x$. From \eqref{12}, conditioned on channel coefficients, this PEP can be expressed as follows:
\begin{align}\label{37}
P(m,x\rightarrow \hat{m},\hat{x})= P\left(  \sum_{l=1}^{n_R}\big| r_l- G_l x \big| ^2  > \sum_{l=1}^{n_R}\big| r_l- \hat{G}_l\hat{x} \big| ^2\right)  
\end{align}
where $G_l=\sum_{i=1}^{N} g_{l,i} e^{j \psi_{m,i}}$ and $\hat{G}_l=\sum_{i=1}^{N} g_{l,i} e^{j \psi_{\hat{m},i}}$. After simple manipulations, we obtain
\begin{align}\label{38}
&P(m,x\rightarrow \hat{m},\hat{x})\nonumber \\
&=P \!\left( \sum_{l=1}^{n_R}\big| G_l  \big| ^2 \big|x \big|^2 \!\!-\! \big| \hat{G}_l  \big| ^2 \big|\hat{x} \big|^2 \!-\!2\Re\!\left\lbrace r_l^* \big( G_lx-\hat{G}_l \hat{x} \big)  \right\rbrace \!> \!0  \right) \nonumber \\
&=P \!\left( \sum_{l=1}^{n_R} -\big| G_l x -\hat{G}_l \hat{x} \big|^2-2\Re\!\left\lbrace n_l^* \big( G_l x -\hat{G}_l \hat{x} \big)  \right\rbrace \!> \!0   \right) \nonumber \\
&=P(G>0)
\end{align}
where we considered the fact that $r_l=G_lx+n_l$ for all $l$. Here, $G\sim\mathcal{N}(\mu_G,\sigma_G^2)$  with $\mu_G= - \sum_{l=1}^{n_R} \big| G_l x -\hat{G}_l \hat{x} \big|^2$  and $\sigma_G^2 =\sum_{l=1}^{n_R} 2N_0\big| G_l x -\hat{G}_l \hat{x} \big|^2 $. Consequently, from $P(G>0)=Q(-\mu_G/ \sigma_G)$, we arrive at
\begin{equation}\label{39}
P(m,x\rightarrow \hat{m},\hat{x}) = Q\left( \sqrt{\frac{\sum_{l=1}^{n_R}\big| G_l x -\hat{G}_l \hat{x} \big|^2}{2N_0}} \right) 
\end{equation}
which is analogous to classical SM-based systems \cite{SM_imperfect}. Defining $\Gamma \triangleq \sum_{l=1}^{n_R}\big| G_l x -\hat{G}_l \hat{x} \big|^2 $ and considering the alternative form of the $Q$-function, the unconditional (averaged over channel coefficients) PEP can be calculated as follows:
\begin{align}\label{40}
&\bar{P}(m,x\rightarrow \hat{m},\hat{x}) = \int_{0}^{\infty} Q\left( \sqrt{\frac{\Gamma}{2N_0}} \right) f_{\Gamma} (\Gamma) d\Gamma \nonumber \\
&\hspace{1cm}=\int_{0}^{\infty} \frac{1}{\pi} \int_{0}^{\pi/2} \exp\left(\frac{-\Gamma}{4\sin^2\eta N_0} \right)  f_{\Gamma} (\Gamma) d\eta d\Gamma \nonumber \\
&\hspace{1cm}= \frac{1}{\pi}\int_{0}^{\pi/2}M_{\Gamma} \left(\frac{-1}{4\sin^2\eta N_0} \right) d\eta.
\end{align}
Here, we need the MGF of $\Gamma$ $(M_{\Gamma}(s))$ to perform this integration. This MGF can be derived by considering the general quadratic form of correlated Gaussian RVs and depends on erroneous or correct detection of the receive antenna index $m$.

\textit{i) First Case} $(m\neq \hat{m})$:  Let us rewrite $\Gamma$ as $\Gamma=\Gamma_1+\Gamma_2+\Gamma_3$, where
\begin{align}\label{41}
\Gamma_1 &=\left| \left[\sum_{i=1}^{N}\beta_{m,i} \right] \!x- \left[\sum_{i=1}^{N} g_{m,i} e^{j \psi_{\hat{m},i}} \right]\! \hat{x}  \right|^2 \!\! = \big| G_m x - \hat{G}_{m} \hat{x} \big|^2 \nonumber \\
\Gamma_2 &= \left| \left[ \sum_{i=1}^{N} g_{\hat{m},i} e^{j \psi_{m,i} }\right]\! x - \left[ \sum_{i=1}^{N} \beta_{\hat{m},i}\right]\! \hat{x}  \right| ^2 \!\!=\big| G_{\hat{m}} x - \hat{G}_{\hat{m}} \hat{x} \big|^2 \nonumber \\
\Gamma_3&=\sum_{l=1  (l\neq m, l \neq \hat{m})}^{n_R}\big| G_l x -\hat{G}_l \hat{x} \big|^2.
\end{align}
Here, $\Gamma_1$, $\Gamma_2$, and $\Gamma_3$ respectively stand for $l=m$, $l=\hat{m}$, and  $l\neq m,l\neq \hat{m}$ in $\Gamma$. Different distributions of $G_l$ and $\hat{G}_l$ with respect to $l$ as well as the correlation among them necessitate the quadratic form of Gaussian RVs to derive $M_{\Gamma}(s)$.

Considering $g_{l,i}=\beta_{l,i} e^{-j\psi_{l,i}}$, let us rewrite $\Gamma_1$ and $\Gamma_2$ as
\begin{align}\label{42}
\Gamma_1 &=\left|\gamma_1 \right| ^2 = (\gamma_1)_{\Re}^2 + (\gamma_1)_{\Im}^2 = \left| \sum_{i=1}^{N} \beta_{m,i} \left(x-e^{-j \bar{\psi_i}} \hat{x}\right) \right|^2 \nonumber \\
\Gamma_2 &= \left|\gamma_2 \right| ^2 = (\gamma_2)_{\Re}^2 + (\gamma_2)_{\Im}^2 = \left| \sum_{i=1}^{N} \beta_{\hat{m},i} \left(xe^{j \bar{\psi_i}} -\hat{x}\right) \right|^2
\end{align}
where $\bar{\psi_i} = \psi_{m,i}-\psi_{\hat{m},i}$ has a triangle-shaped PDF defined in \eqref{14}. It is obvious from \eqref{42} and the CLT that $\gamma_1$ and $\gamma_2$ follow complex Gaussian distribution for increasing $N$, however, we need to consider the correlation among their components. After tedious but straightforward calculations, the mean vector and the covariance matrix of $\mathbf{g}=\begin{bmatrix}
(\gamma_1)_{\Re} & (\gamma_1)_{\Im} & (\gamma_2)_{\Re} & (\gamma_2)_{\Im}
\end{bmatrix}^{\mathrm{T}}$ are obtained respectively as follows:
\begin{equation}\label{43}
\mathbf{m}=\begin{bmatrix}
\frac{N\sqrt{\pi}x_{\Re}}{2} & \frac{N\sqrt{\pi}x_{\Im}}{2} & -\frac{N\sqrt{\pi}\hat{x}_{\Re}}{2} & -\frac{N\sqrt{\pi}\hat{x}_{\Im}}{2}
\end{bmatrix}^\mathrm{T}
\end{equation}
\begin{equation}\label{44}
\mathbf{C}=\!\begin{bmatrix}
\sigma_1^2  & \sigma_{1,2} & \sigma_{1,3} & \sigma_{1,4} \\
\sigma_{1,2} &\sigma_2^2 & \sigma_{2,3} & \sigma_{2,4} \\
\sigma_{1,3} & \sigma_{2,3} & \sigma_3^2& \sigma_{3,4} \\
\sigma_{1,4} & \sigma_{2,4} & \sigma_{3,4} & \sigma_4^2
\end{bmatrix}
\end{equation}
where
\begin{align*}
\sigma_1^2 & = \tfrac{N(4-\pi)x_{\Re}^2 }{4} \!+\! \tfrac{N\left|\hat{x} \right|^2 }{2}, \,\,\, \sigma_2^2  =\tfrac{N(4-\pi)x_{\Im}^2 }{4} \!+\! \tfrac{N\left|\hat{x} \right|^2 }{2} \\
\sigma_3^2 & =\tfrac{N(4-\pi)\hat{x}_{\Re}^2 }{4} \!+\! \tfrac{N\left|x \right|^2 }{2}, \,\,\, \sigma_4^2  =\tfrac{N(4-\pi)\hat{x}_{\Im}^2 }{4} \!+\! \tfrac{N\left|x \right|^2 }{2} \\
\sigma_{1,2}&= \tfrac{N(4-\pi)x_{\Re} x_{\Im}}{4}  , \,\,\, \sigma_{3,4}=  \tfrac{N(4-\pi)\hat{x}_{\Re} \hat{x}_{\Im}}{4}  \\
\sigma_{1,3}&=\tfrac{N\pi (-x_{\Re} \hat{x}_{\Re} + x_{\Im} \hat{x}_{\Im})}{8}  , \,\,\, \sigma_{1,4}=-\tfrac{N\pi (x_{\Re} \hat{x}_{\Im} + \hat{x}_{\Re} x_{\Im}) }{8}  \\
\sigma_{2,3}&=\sigma_{1,4}, \,\,\, \sigma_{2,4}=-\sigma_{1,3}.
\end{align*}
Substituting \eqref{43} and \eqref{44} in the MGF of the quadratic form of Gaussian RVs given in \eqref{31} for $\mathbf{A}=\mathbf{I}_4$, yields the MGF of $\Gamma_1+\Gamma_2=\mathbf{g}^{\mathrm{T}} \mathbf{A} \mathbf{g}$. On the other hand, $\Gamma_3$ can be rewritten as
\begin{equation}\label{45}
\Gamma_3=\sum_{l=1 (l\neq m, l \neq \hat{m})}^{n_R}\left| \sum_{i=1}^{N} g_{l,i} \left( xe^{j \psi_{m,i}} - \hat{x}e^{j \psi_{\hat{m},i}} \right)  \right|^2.
\end{equation}
Fortunately, due to zero means of $g_{l,i}$ and $x_i= xe^{j \psi_{m,i}} - \hat{x}e^{j \psi_{\hat{m},i}}$, and their independence for all $l\, (l\neq m, l \neq \hat{m})$ and $i$, we have $\mathrm{Var}[(g_{l,i}x_i)_{\Re}]=\mathrm{Var}[(g_{l,i}x_i)_{\Im}]=\frac{\left|x \right|^2 + \left|\hat{x} \right|^2 }{2}$. Consequently, for large $N$, $\Gamma_3$ can be expressed as the sum of $n_R-2$ independent central $\chi^2$ RVs with two degrees of freedom, and has the following simple MGF:
\begin{equation}\label{46}
M_{\Gamma_3}(s)=\left(\frac{1}{1-sN(\left|x \right|^2 + \left|\hat{x} \right|^2)} \right) ^{n_R-2}.
\end{equation}   
Finally, substituting the MGF of $\Gamma$, obtained from the product of MGFs of $\Gamma_1+\Gamma_2$ and $\Gamma_3$, in \eqref{40} and evaluating this simple integral numerically yields the desired unconditional PEP. It is worth noting that the unconditional PEP is independent of $m$ and $\hat{m}$.

\textit{ii) Second Case} $(m= \hat{m})$: For the calculation of PEP in case of correctly detected receive antenna indices, considering $G_l=\hat{G}_l$, we can rewrite $\Gamma$ as
\begin{equation}\label{47}
\Gamma= \sum_{l=1}^{n_R} \big|G_l(x-\hat{x}) \big|^2 =\big|x-\hat{x} \big|^2 \left( G_m^2 + \sum_{l=1 (l\neq m)}^{n_R} \left| G_l \right|^2 \right). 
\end{equation}
Keeping in mind that $G_m\sim \mathcal{N}(N\sqrt{\pi}/2,N(4-\pi)/4)$ and $G_l \sim \mathcal{CN}(0,N) $ for $l\neq m$ (see Section III.A), we obtain 
\begin{align}\label{48}
M_{\Gamma}(s)&=\left( \frac{1}{1-\frac{sN(4-\pi)\left|x-\hat{x} \right|^2}{2}}\right)^{\!\frac{1}{2}} \exp \left(\frac{ \frac{sN^2 \left|x-\hat{x} \right|^2 \pi}{4}}{1-\frac{sN(4-\pi)\left|x-\hat{x} \right|^2}{2}} \right) \nonumber  \\
&\hspace{1cm} \times \left( \frac{1}{1-sN\left|x-\hat{x} \right|^2}\right)^{n_R-1}.  
\end{align}
Substituting this MGF in \eqref{40} and performing numerical integration provides the desired PEP.

Finally, the PEP values obtained from \eqref{40} for both cases are used to derive the following union bound on BEP:
\begin{equation}\label{49}
P_b \!\le\! \frac{1}{Mn_R} \sum_{m} \sum_{\hat{m}} \sum_{x} \sum_{\hat{x}} \frac{\bar{P}(m,x\rightarrow \hat{m},\hat{x}) e(m,x\rightarrow \hat{m},\hat{x})}{\log_2(Mn_R)},
\end{equation}
where $e(m,x\rightarrow \hat{m},\hat{x})$ stands for the number of bits in error for the corresponding pairwise error event. 

\textit{Remark 3}: The above analysis is general and can be considered for all constellations. Derivation of  simplified expressions for BPSK and QPSK (or $M$-PSK in general) is left to interested readers.

  \begin{figure}[!t]
	\begin{center}
		\includegraphics[width=0.85\columnwidth]{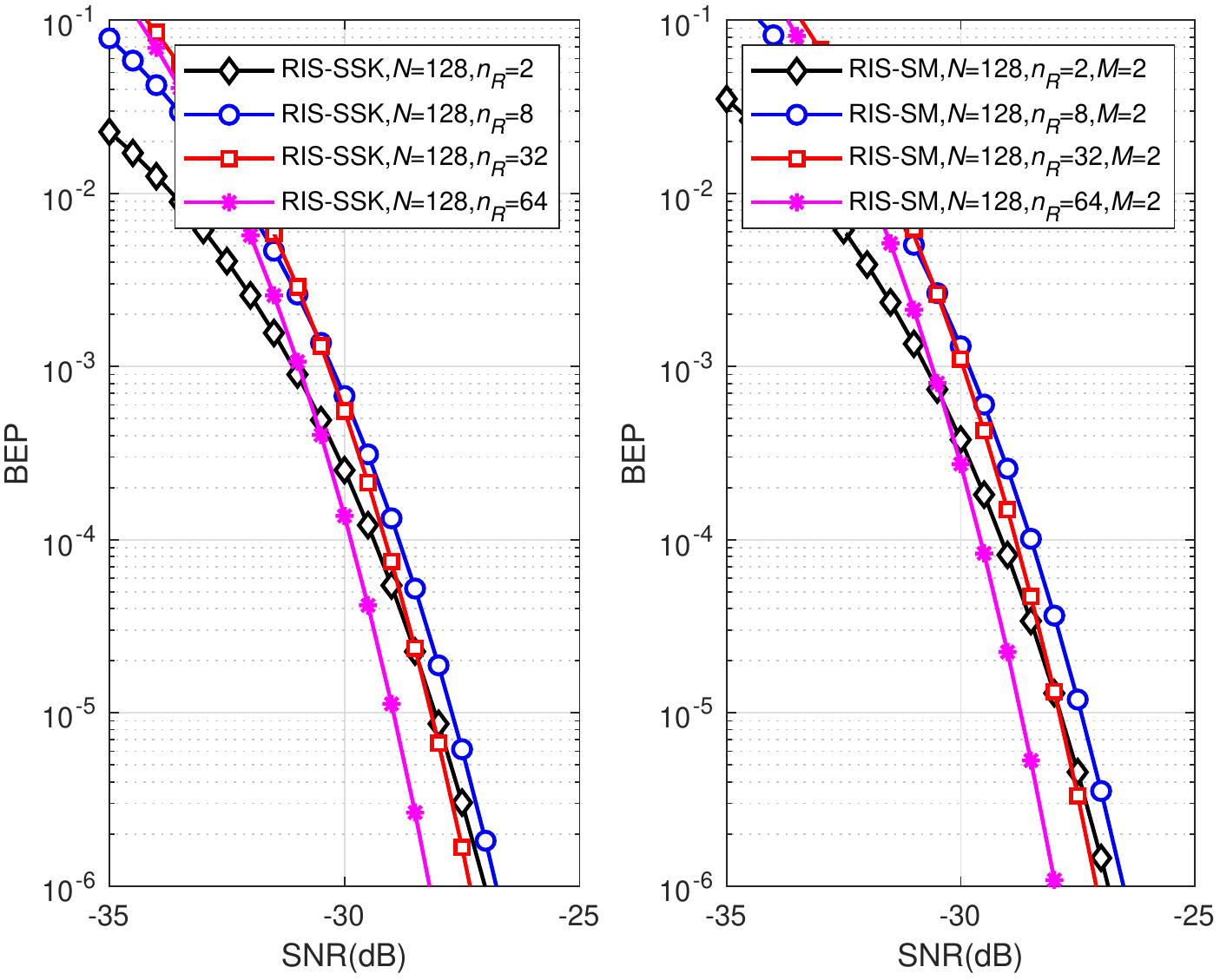}
		\vspace*{-0.4cm}\caption{Theoretical BEP performance of RIS-SSK and RIS-SM systems with increasing $n_R$.}\vspace*{-0.4cm}
		\label{Fig6}
	\end{center}
\end{figure}

\vspace*{-0.35cm}

\subsection{Performance of RIS-SSK}
Considering $x=\hat{x}=\sqrt{E_s}$ (i.e., an unmodulated carrier in the baseband) in \eqref{39}, we obtain the conditional PEP of the RIS-SSK scheme as
\begin{equation}\label{50}
P(m\rightarrow \hat{m}) = Q\left( \sqrt{\frac{\sum_{l=1}^{n_R}E_s\big| G_l  -\hat{G}_l  \big|^2}{2N_0}} \right). 
\end{equation}
In light of this information, the analyses in Section IV.A (for the case of $m \neq \hat{m}$) is also valid for ML detection of RIS-SSK and the unconditional PEP $\bar{P}(m\rightarrow \hat{m})$ can be easily derived from \eqref{40} with suitable modifications in $M_{\Gamma}(s)$. Then the BEP upper bound of RIS-SSK can be calculated similar to \eqref{23} as
 \begin{equation}\label{51}
 P_b  \le \frac{n_R}{2}  \bar{P}(m \rightarrow \hat{m}).
 \end{equation}

\textit{Remark 4}: We observe that unlike the greedy detector, increasing $n_R$ for ML detection improves the PEP performance of RIS-SM and RIS-SSK schemes through the MGF terms of \eqref{46} and \eqref{48}, which  include $n_R$ in their exponents. Since increasing $n_R$ also improves the data rate along with increasing number of bit errors in \eqref{49} and \eqref{51}, we face an interesting trade-off among complexity, performance, and data rate. In Fig. \ref{Fig6}, we show the theoretical BEP performance of RIS-SSK and RIS-SM schemes calculated from \eqref{49} and \eqref{51} for $N=128$ reflectors and increasing $n_R$, with respect to $E_s/N_0$. As seen from Fig. \ref{Fig6}, increasing $n_R$ eventually improves the BER performance (since $n_R$ appears in \eqref{51}), while providing a higher data rate, which is quite unusual for legacy communication systems not exploiting IM. In other words, increasing $n_R$ both improves the spectral efficiency and overall BER performance for RIS-SSK/SM systems in the presence of ML detection.   
 
\begin{figure}[!t]
	\begin{center}
		\includegraphics[width=0.85\columnwidth]{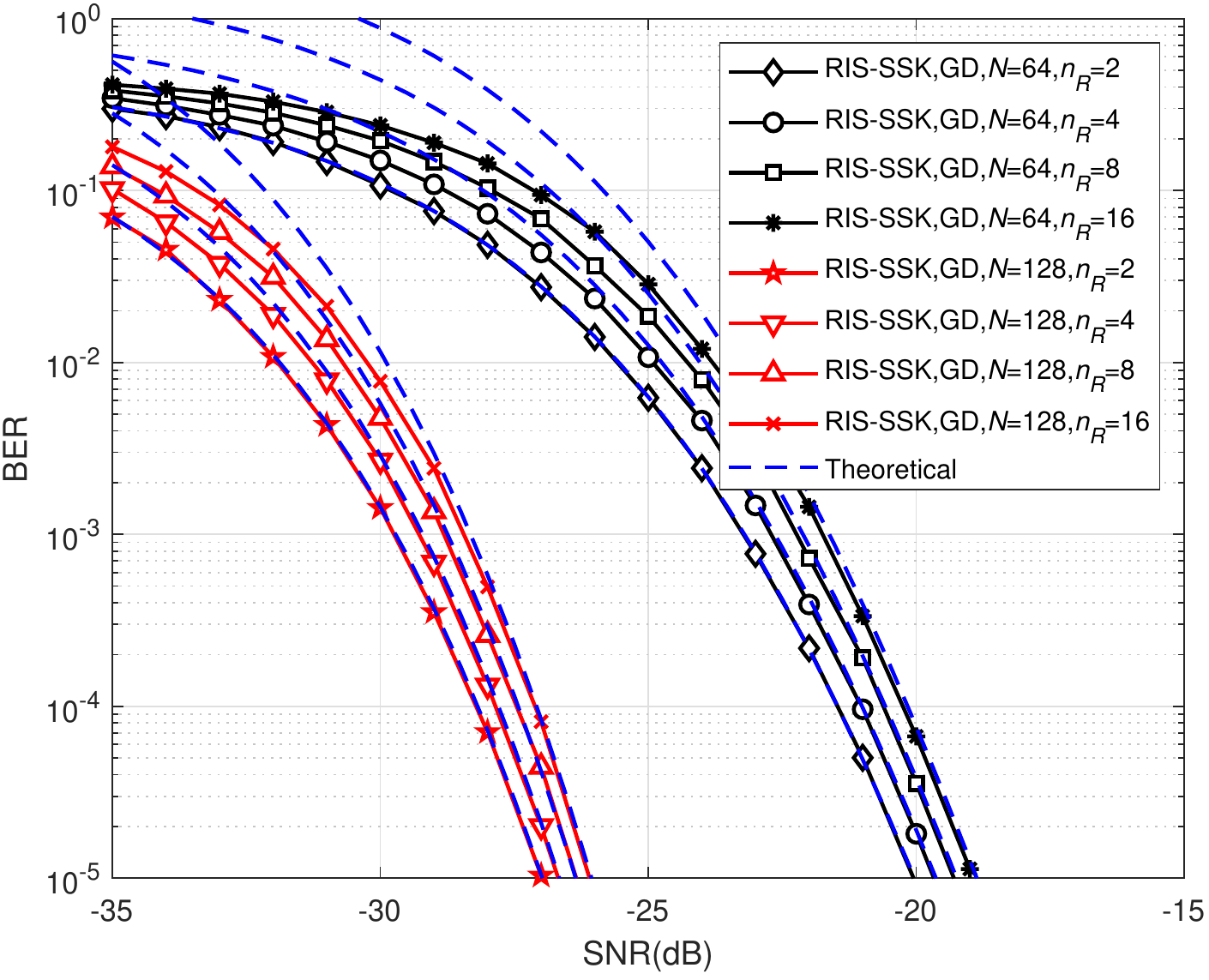}
		\vspace*{-0.4cm}\caption{Theoretical and computer simulation results for RIS-SSK with greedy detection.}\vspace*{-0.4cm}
		\label{Fig1}
	\end{center}
\end{figure}

\section{Simulation Results}
In this section, we provide computer simulation results for the proposed RIS-based SSK and SM schemes and make comparisons with our theoretical results and reference schemes. We consider $E_s/N_0$ as the SNR, similar to the classical diversity combining and space modulation schemes. For clarity, as usual practice, the
large-scale path-loss is not considered since it is implicitly taken into account in the received
SNR.

In Figs. 4 and 5, we provide BER performance curves of the RIS-SSK and RIS-SM systems for greedy detection and make comparisons with the theoretical results obtained from \eqref{23} and \eqref{24}, respectively. As seen from the given results, our theoretical findings are quite accurate for both schemes and the performance of RIS-SSK and RIS-SM schemes degrade with increasing bits per channel use (bpcu), or equivalently $n_R$, values (see Remarks 1 and 2). It is worth noting that BER performance of both schemes significantly improves by increasing the $N$ value from $64$ to $128$, which is consistent with \eqref{22}.

\begin{figure}[!t]
	\begin{center}
		\includegraphics[width=0.85\columnwidth]{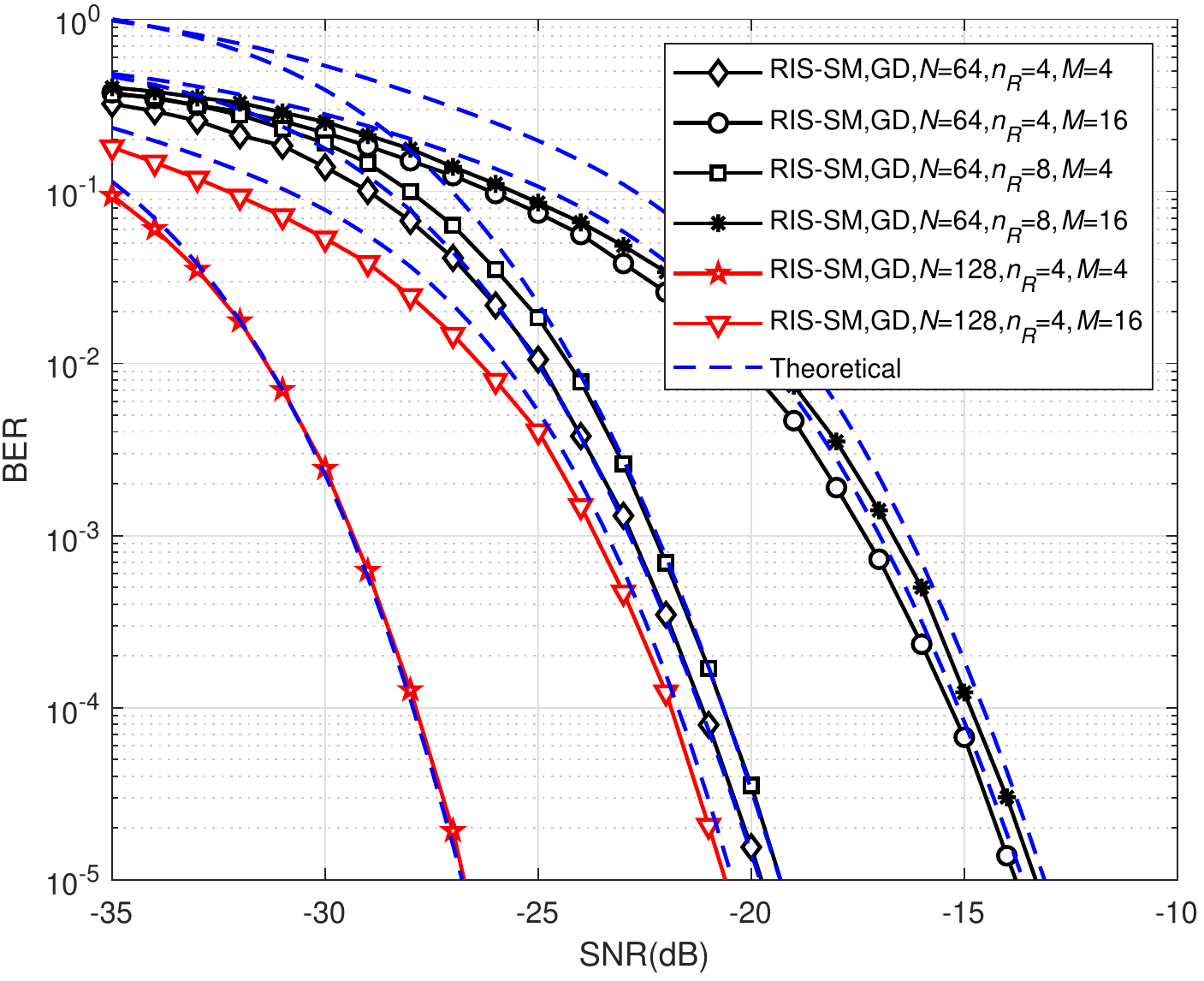}
		\vspace*{-0.4cm}\caption{Theoretical and computer simulation results for RIS-SM with greedy detection.}\vspace*{-0.4cm}
		\label{Fig2}
	\end{center}
\end{figure}

\begin{figure}[!t]
	\begin{center}
		\includegraphics[width=0.85\columnwidth]{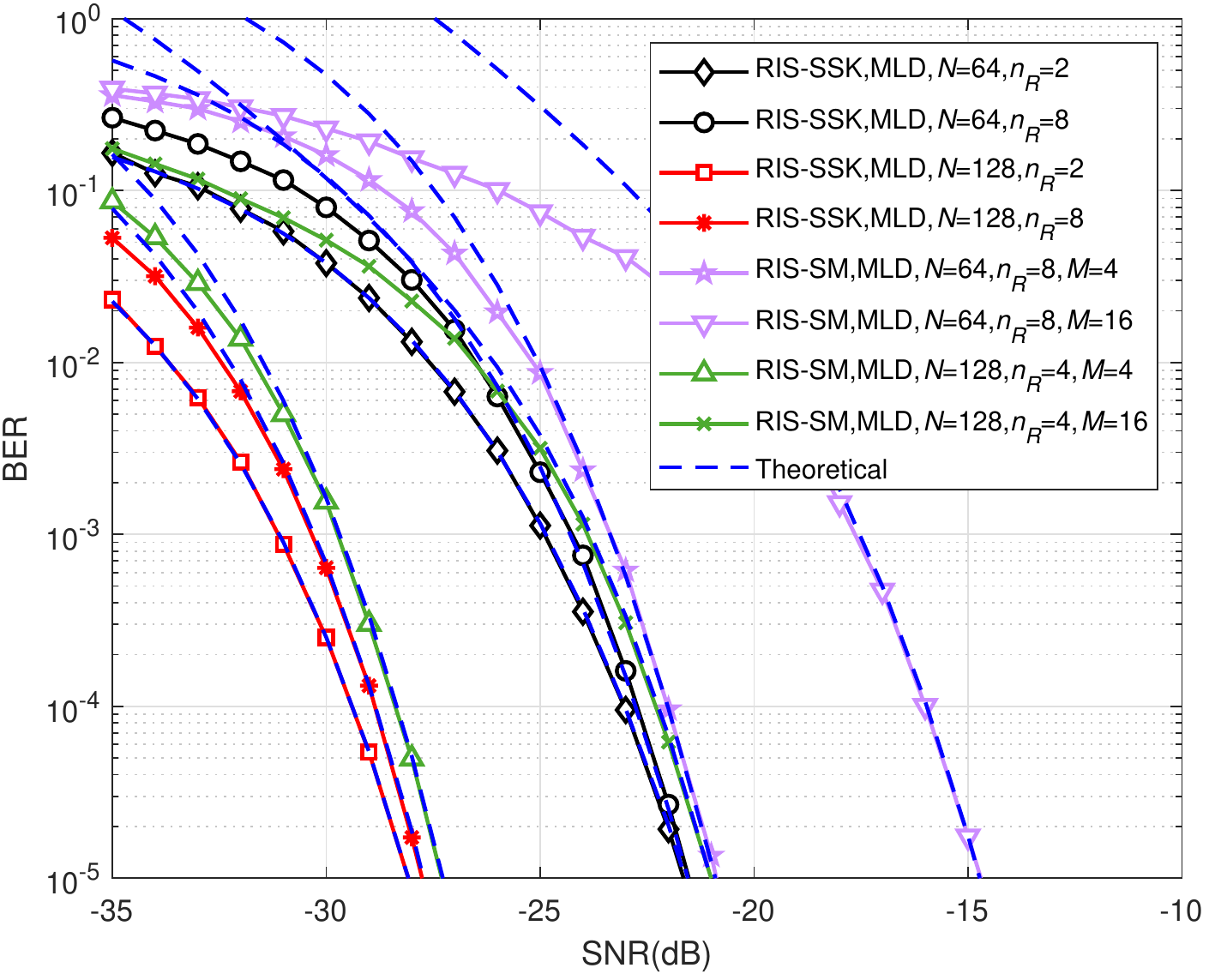}
		\vspace*{-0.4cm}\caption{Theoretical and computer simulation results for RIS-SSK and RIS-SM with ML detection.}\vspace*{-0.4cm}
		\label{Fig3}
	\end{center}
\end{figure}

\begin{figure}[!t]
	\begin{center}
		\includegraphics[width=0.85\columnwidth]{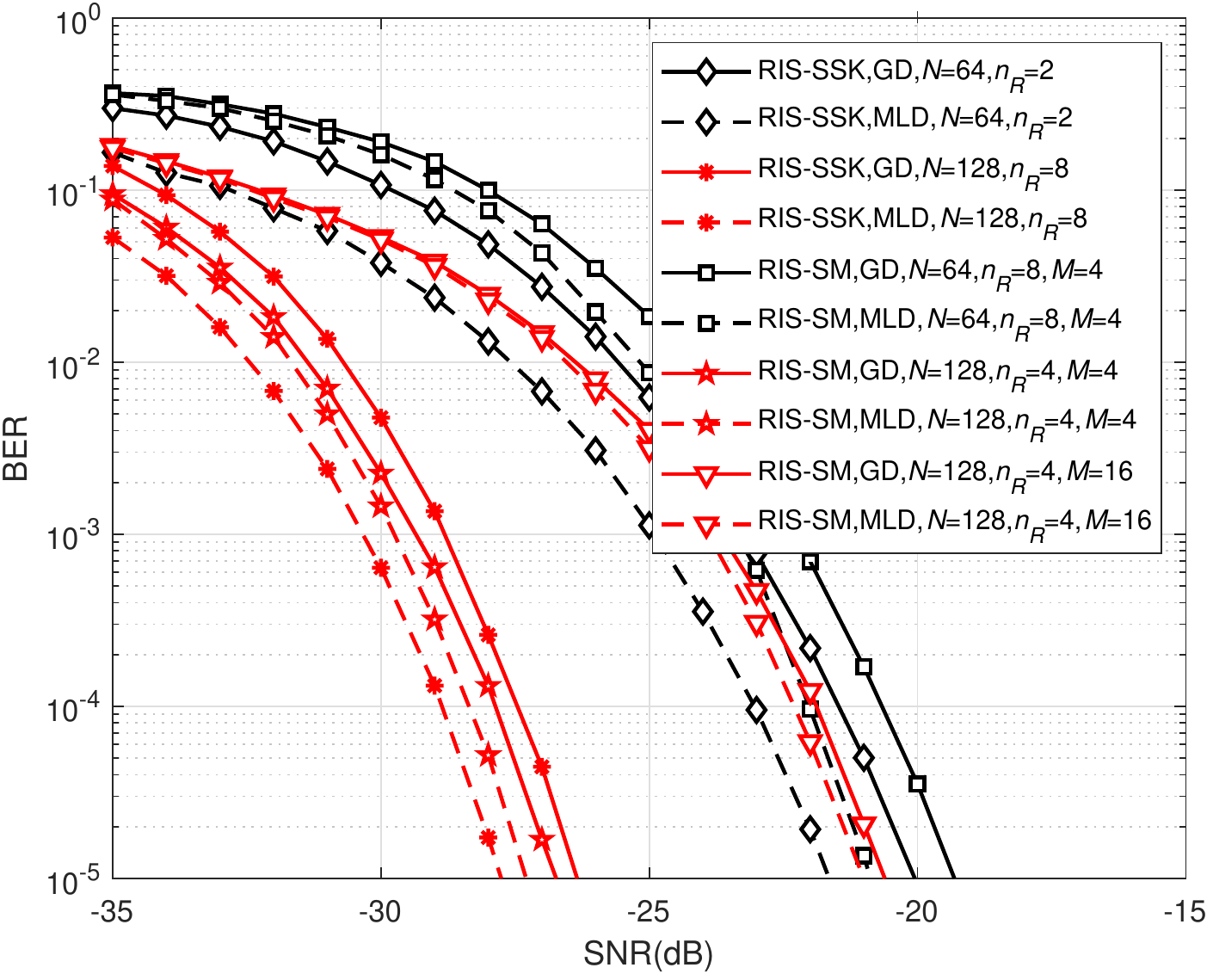}
		\vspace*{-0.4cm}\caption{BER performance comparisons for greedy and ML detectors of RIS-SSK and RIS-SM schemes.}\vspace*{-0.4cm}
		\label{Fig4}
	\end{center}
\end{figure}

In Fig. 6, we focus on the performance of RIS-SSK and RIS-SM schemes with ML detection and make comparisons with the theoretical results obtained from \eqref{49} and \eqref{51}. As seen from Fig. 6, while increasing $n_R$ does not cause a remarkable BER degradation for the RIS-SSK scheme, the effect of increasing $M$ is more evident for the RIS-SM scheme. It is worth noting that the performance of the ML detector improves with $  N $ as well due to (\ref{46}) and (\ref{48}).

We compare the BER performances of greedy and ML detectors for both RIS-SSK and RIS-SM systems at various bpcu values in Fig. 7. We observe that the ML detector of RIS-SSK provides approximately $2$ dB improvement in the required SNR for the considered two setups ($N=64,n_R=2$ and $N=128,n_R=8$). Although we observe a similar improvement for RIS-SM in case of $N=64$, the difference in BER performances of greedy and ML detectors is relatively smaller for the case of $N=128$. The gap between the greedy and ML detectors of RIS-SSK can be attributed to its nature, which considers only indices to convey information.

Finally, in Fig. 8, we present BER performance comparison results for RIS-SSK, RIS-SM, RIS-AP \cite{Basar_2019_LIS}, and conventional fully-digital (zero-forcing) precoding-based receive SSK (RSSK) \cite{Zhang_2013} at $3,4,$ and $6$ bpcu values with ML detection. We have several important observations from Fig. 8. First, an interesting trade-off exists between the receiver cost and the BER performance for RIS-SSK and RIS-SM schemes: while the former provides a better BER performance, the latter exhibits a slight degradation by using a less number of receive antennas at the same bpcu. Second, the RIS-AP scheme proposed in \cite{Basar_2019_LIS}, which utilizes a single receive antenna, cannot compete with RIS-SSK/SM schemes since it creates a virtual $M$-PSK constellation by altering RIS phases and suffers at high bpcu values. Third, compared to RIS-based new schemes, a more than $15$ dB difference in required SNR  is observed for the conventional RSSK-MIMO scheme. Although utilizing a massive MIMO system, RSSK forces the MIMO channels into zero to realize a pure SSK-like reception, while RIS-based SM/SSK schemes constructively exploit the wireless channels to boost the signal quality at the intended receive antenna. In other words, RIS-based schemes exploit the inherent randomness of fading channels in a more effective way by the constructive alignment of reflected signals. However, although both solutions have similar level of algorithmic complexity, RIS-based schemes may have higher signaling overhead due to the training of RISs, while traditional schemes have a much higher transmitter complexity.  

\begin{figure}[t]
	\begin{center}
		\includegraphics[width=0.84\columnwidth]{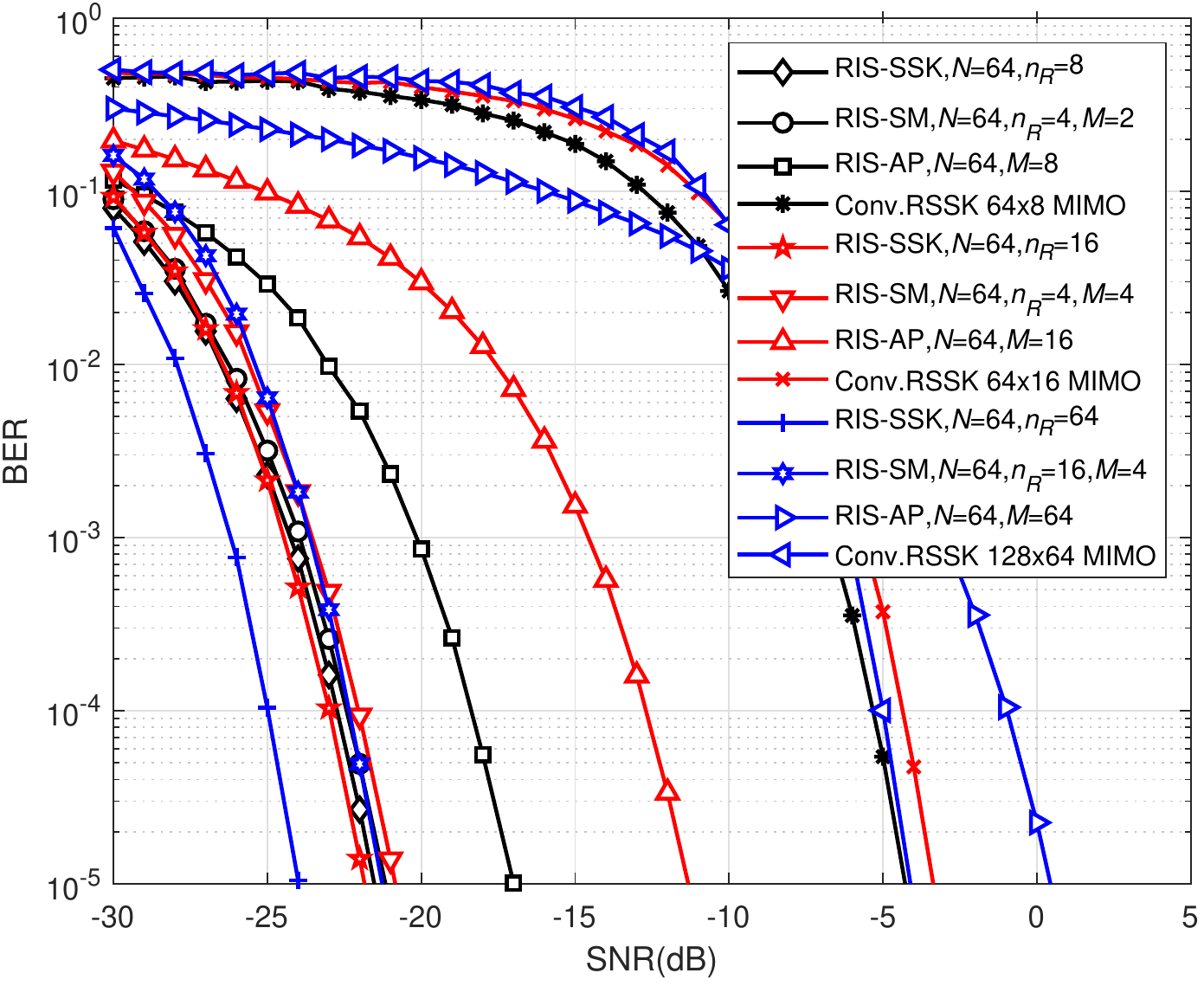}
		\vspace*{-0.4cm}\caption{BER performance comparison of RIS-SSK, RIS-SM, RIS-AP \cite{Basar_2019_LIS}, and conventional RSSK \cite{Zhang_2013} schemes for $3$, $4$, and $6$ bpcu.}\vspace*{-0.4cm}
		\label{Fig5}
	\end{center}
\end{figure}

\begin{figure}[t]
	\begin{center}
		\includegraphics[width=0.85\columnwidth]{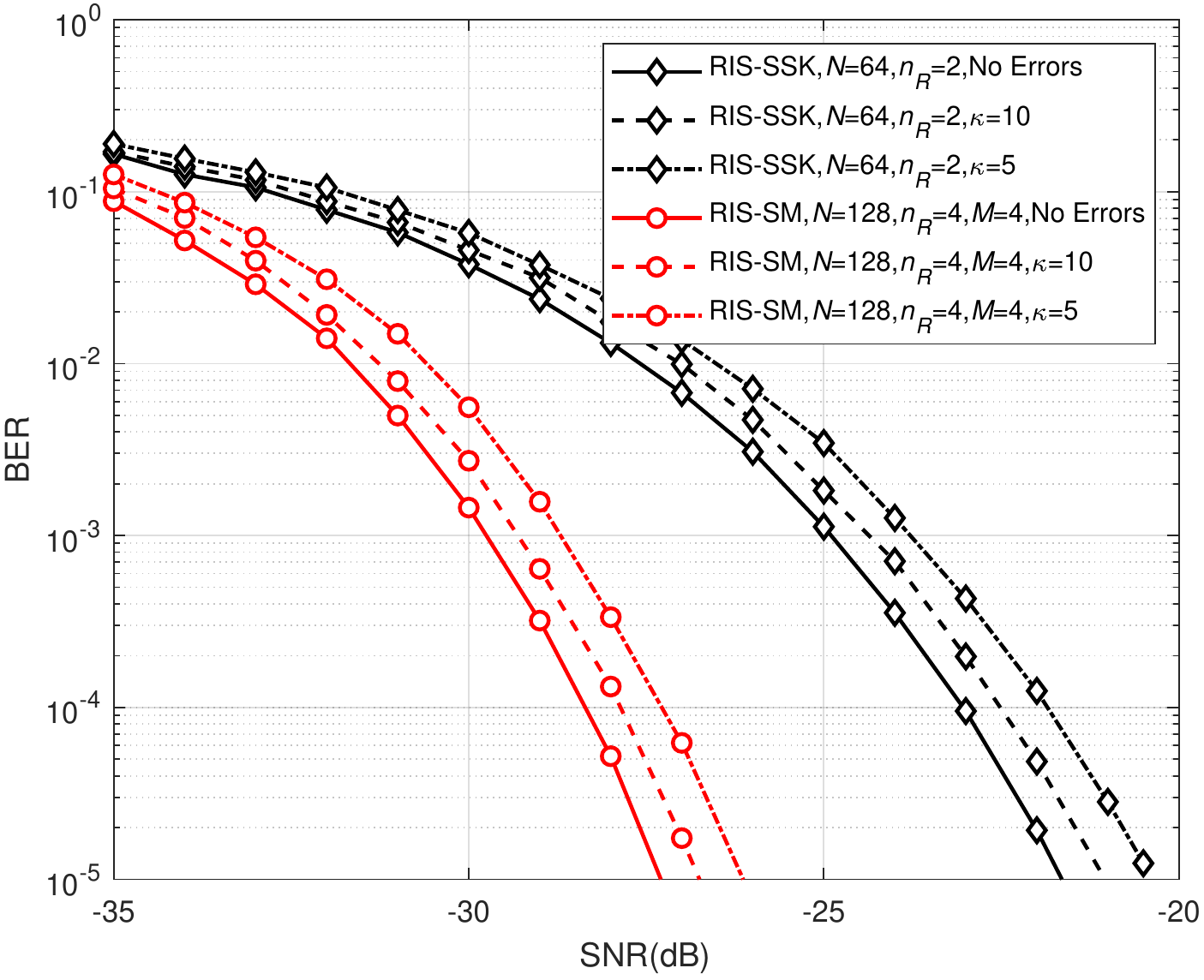}
		\vspace*{-0.45cm}\caption{Performance of RIS-SSK and RIS-SM schemes in the presence of channel phase estimation errors.}\vspace*{-0.4cm}
		\label{Fig_new}
	\end{center}
\end{figure}

In Fig. \ref{Fig_new}, we investigate the performance of proposed RIS-SSK and RIS-SM schemes with erroneous channel phase estimates at the RIS. Here, the phase estimation error at the RIS is modelled according to von Mises (circular normal) distribution, whose concentration parameter $\kappa$ captures the accuracy of the estimation \cite{Coon_2019}. As seen from Fig. \ref{Fig_new}, while the degradation in BER is not significant for $\kappa=10$ (higher accuracy estimation), a noticeable difference is observed for $\kappa=5$ for both proposed schemes.

\section{Conclusions}
The general concept of RIS-assisted IM has been proposed in this paper as a new beyond massive MIMO paradigm for next-generation (potentially 6G or beyond) wireless networks. It has been shown by comprehensive theoretical derivations as well as computer simulations that the proposed RIS-SSK and RIS-SM schemes have the potential to provide considerably high spectral efficiency at extremely low SNR values through a smart and RIS-assisted indexing mechanism for available receive antennas. We conclude that the effective use of RIS-assisted communication schemes may be a game-changing paradigm for next-generation communication networks by eliminating the need for sophisticated massive MIMO schemes that require expensive and power-hungry components. The extremely low SNR regions of operation may also be a remedy to the increasing need for advanced channel coding schemes to achieve ultra-reliable communications. As also discussed in Introduction, potential applications of IM for transmit antennas and/or RIS regions along with other advanced/generalized schemes, the design of low-complexity receiver architectures, and analyses in the presence of potential system imperfections and more sophisticated channel/correlation models, remain as interesting and open research problems. We also hope that the theoretical insights provided in this work may be useful for future implementation and hardware impairment related challenges in the context of emerging RISs.

\bibliographystyle{IEEEtran}
\bibliography{IEEEabrv,bib_2020}

\begin{thebibliography}{10}
\providecommand{\url}[1]{#1}
\csname url@samestyle\endcsname
\providecommand{\newblock}{\relax}
\providecommand{\bibinfo}[2]{#2}
\providecommand{\BIBentrySTDinterwordspacing}{\spaceskip=0pt\relax}
\providecommand{\BIBentryALTinterwordstretchfactor}{4}
\providecommand{\BIBentryALTinterwordspacing}{\spaceskip=\fontdimen2\font plus
\BIBentryALTinterwordstretchfactor\fontdimen3\font minus
  \fontdimen4\font\relax}
\providecommand{\BIBforeignlanguage}[2]{{%
\expandafter\ifx\csname l@#1\endcsname\relax
\typeout{** WARNING: IEEEtran.bst: No hyphenation pattern has been}%
\typeout{** loaded for the language `#1'. Using the pattern for}%
\typeout{** the default language instead.}%
\else
\language=\csname l@#1\endcsname
\fi
#2}}
\providecommand{\BIBdecl}{\relax}
\BIBdecl

\bibitem{6G}
\BIBentryALTinterwordspacing
A.~Gatherer. (2018, June) What will 6{G} be? [Online]. Available:
  \url{https://www.comsoc.org/publications/ctn/what-will-6g-be}
\BIBentrySTDinterwordspacing

\bibitem{Saad_2019}
W.~{Saad}, M.~{Bennis}, and M.~{Chen}, ``A vision of 6{G} wireless systems:
  Applications, trends, technologies, and open research problems (to appear),''
  \emph{IEEE Netw.}, pp. 1--9, Oct. 2019.

\bibitem{Khandani_conf1}
E.~Seifi, M.~Atamanesh, and A.~K. Khandani, ``Media-based {MIMO}:
  {O}utperforming known limits in wireless,'' in \emph{Proc. 2016 IEEE Int.
  Conf. Commun. (ICC)}, Kuala Lumpur, Malaysia, May 2016, pp. 1--7.

\bibitem{MBM_TVT}
Y.~Naresh and A.~Chockalingam, ``On media-based modulation using {RF}
  mirrors,'' \emph{IEEE Trans. Veh. Technol.}, vol.~66, no.~6, pp. 4967--4983,
  June 2017.

\bibitem{Basar_2019}
E.~Basar, ``Media-based modulation for future wireless systems: {A} tutorial,''
  \emph{IEEE Wireless Commun.}, vol.~26, no.~5, pp. 160--166, Oct. 2019.

\bibitem{SSM}
Y.~{Ding}, K.~J. {Kim}, T.~{Koike-Akino}, M.~{Pajovic}, P.~{Wang}, and
  P.~{Orlik}, ``Spatial scattering modulation for uplink millimeter-wave
  systems,'' \emph{IEEE Commun. Lett.}, vol.~21, no.~7, pp. 1493--1496, July
  2017.

\bibitem{BIM}
Y.~{Ding}, V.~{Fusco}, A.~{Shitvov}, Y.~{Xiao}, and H.~{Li}, ``Beam index
  modulation wireless communication with analog beamforming,'' \emph{IEEE
  Trans. Veh. Technol.}, vol.~67, no.~7, pp. 6340--6354, July 2018.

\bibitem{Basar_2017}
E.~Basar, M.~Wen, R.~Mesleh, M.~D. Renzo, Y.~Xiao, and H.~Haas, ``Index
  modulation techniques for next-generation wireless networks,'' \emph{IEEE
  Access}, vol.~5, pp. 16\,693--16\,746, Sept. 2017.

\bibitem{Akyildiz_2018}
C.~{Liaskos}, S.~{Nie}, A.~{Tsioliaridou}, A.~{Pitsillides}, S.~{Ioannidis},
  and I.~{Akyildiz}, ``A new wireless communication paradigm through
  software-controlled metasurfaces,'' \emph{IEEE Commun. Mag.}, vol.~56, no.~9,
  pp. 162--169, Sept. 2018.

\bibitem{Di_Renzo_2019}
M.~{Di Renzo \it{et al.}}, ``Smart radio environments empowered by
  reconfigurable {AI} meta-surfaces: {A}n idea whose time has come,''
  \emph{EURASIP J. Wireless Commun. Net.}, vol. 2019, p. 129, May 2019.

\bibitem{Basar_Access_2019}
E.~Basar, M.~D. Renzo, J.~de~Rosny, M.~Debbah, M.-S. Alouini, and R.~Zhang,
  ``Wireless communications through reconfigurable intelligent surfaces,''
  \emph{IEEE Access}, p. 116753–116773, Sep. 2019.

\bibitem{Subrt_2012}
L.~{Subrt} and P.~{Pechac}, ``Controlling propagation environments using
  intelligent walls,'' in \emph{Proc. 2012 6th European Conf. Antennas Propag.
  ({EUCAP})}, Prague, Czech Republic, Mar. 2012, pp. 1--5.

\bibitem{Tan_2016}
X.~{Tan}, Z.~{Sun}, J.~M. {Jornet}, and D.~{Pados}, ``Increasing indoor
  spectrum sharing capacity using smart reflect-array,'' in \emph{Proc. 2016
  IEEE Int. Conf. Commun. (ICC)}, Kuala Lumpur, Malaysia, May 2016, pp. 1--6.

\bibitem{Huang_2018}
C.~{Huang}, A.~{Zappone}, M.~{Debbah}, and C.~{Yuen}, ``Achievable rate
  maximization by passive intelligent mirrors,'' in \emph{Proc. 2018 IEEE Int.
  Conf. Acoust. Speech Signal Process. (ICASSP)}, Calgary, Canada, Apr. 2018,
  pp. 3714--3718.

\bibitem{Huang_2018_2}
C.~{Huang}, G.~C. {Alexandropoulos}, A.~{Zappone}, M.~{Debbah}, and C.~{Yuen},
  ``Energy efficient multi-user {MISO} communication using low resolution large
  intelligent surfaces,'' in \emph{Proc. IEEE Global Commun. Conf.}, Abu Dhabi,
  UAE, Dec. 2018.

\bibitem{Huang_2019}
C.~{Huang}, A.~{Zappone}, G.~C. {Alexandropoulos}, M.~{Debbah}, and C.~{Yuen},
  ``Reconfigurable intelligent surfaces for energy efficiency in wireless
  communication,'' \emph{IEEE Trans. Wireless Commun.}, vol.~8, Aug. 2019.

\bibitem{Wu_2018}
Q.~Wu and R.~Zhang, ``Intelligent reflecting surface enhanced wireless network:
  {J}oint active and passive beamforming design,'' in \emph{Proc. IEEE Global
  Commun. Conf.}, Abu Dhabi, UAE, Dec. 2018.

\bibitem{Wu_2018_2}
\BIBentryALTinterwordspacing
------, ``Beamforming optimization for intelligent reflecting surface with
  discrete phase shifts,'' in \emph{Proc. 2019 IEEE Int. Conf. Acoust. Speech
  Signal Process. (ICASSP)}, Brighton, UK, May 2019. [Online]. Available:
  \url{arXiv:1810.10718}
\BIBentrySTDinterwordspacing

\bibitem{Taha_2019}
\BIBentryALTinterwordspacing
A.~Taha, M.~Alrabeiah, and A.~Alkhateeb, ``Enabling large intelligent surfaces
  with compressive sensing and deep learning,'' Apr. 2019. [Online]. Available:
  \url{arXiv:1904.10136}
\BIBentrySTDinterwordspacing

\bibitem{Schober_2019_2}
\BIBentryALTinterwordspacing
X.~Yu, D.~Xu, and R.~Schober, ``Enabling secure wireless communications via
  intelligent reflecting surfaces,'' in \emph{Proc. IEEE Global Commun. Conf.
  (GLOBECOM)}, Waikoloa, HI, USA, Apr. 2019. [Online]. Available:
  \url{arXiv:1904.09573}
\BIBentrySTDinterwordspacing

\bibitem{Basar_2019_LIS}
E.~Basar, ``Transmission through large intelligent surfaces: {A} new frontier
  in wireless communications,'' in \emph{Proc. European Conf. Netw. Commun.
  (EuCNC 2019)}, Valencia, Spain, June 2019.

\bibitem{Wu_2019}
\BIBentryALTinterwordspacing
Q.~Wu and R.~Zhang, ``Towards smart and reconfigurable environment: Intelligent
  reflecting surface aided wireless network,'' \emph{IEEE Commun. Mag. (to
  appear)}. [Online]. Available: \url{arXiv:1905.00152v2}
\BIBentrySTDinterwordspacing

\bibitem{IM_5G}
E.~Basar, ``Index modulation techniques for 5{G} wireless networks,''
  \emph{IEEE Commun. Mag.}, vol.~54, no.~7, pp. 168--175, June 2016.

\bibitem{IM_Book}
M.~Wen, X.~Cheng, and L.~Yang, \emph{Index Modulation for 5G Wireless
  Communications}.\hskip 1em plus 0.5em minus 0.4em\relax Berlin, Germany:
  Springer, 2017.

\bibitem{Mesleh_2018}
R.~Mesleh and A.~Alhassi, \emph{Space Modulation Techniques}.\hskip 1em plus
  0.5em minus 0.4em\relax New Jersey: John Wiley \& Sons, 2018.

\bibitem{OFDM_IM}
E.~Basar, U.~Aygolu, E.~Panayirci, and H.~V. Poor, ``Orthogonal frequency
  division multiplexing with index modulation,'' \emph{{IEEE} Trans. Signal
  Process.}, vol.~61, no.~22, pp. 5536--5549, Nov. 2013.

\bibitem{Abeywickrama_2019}
\BIBentryALTinterwordspacing
S.~Abeywickrama, R.~Zhang, and C.~Yuen, ``Intelligent reflecting surface:
  Practical phase shift model and beamforming optimization,'' July 2019.
  [Online]. Available: \url{arXiv:1907.06002}
\BIBentrySTDinterwordspacing

\bibitem{He_2019}
\BIBentryALTinterwordspacing
Z.-Q. He and X.~Yuan, ``Cascaded channel estimation for large intelligent
  metasurface assisted massive {MIMO},'' May 2019. [Online]. Available:
  \url{arXiv:1905.07948}
\BIBentrySTDinterwordspacing

\bibitem{Simon2}
M.~Simon, \emph{Probability Distributions Involving {G}aussian Random
  Variables}.\hskip 1em plus 0.5em minus 0.4em\relax New York: Springer, 2002.

\bibitem{Mathai_1992}
A.~Mathai and S.~B. Provost, \emph{Quadratic Forms in Random Variables:
  {T}heory and Applications}.\hskip 1em plus 0.5em minus 0.4em\relax New York:
  Marcel Dekker, 1992.

\bibitem{Simon}
M.~Simon and M.~S. Alouini, \emph{Digital Communications over Fading Channels},
  2nd~ed.\hskip 1em plus 0.5em minus 0.4em\relax New Jersey: John Wiley \&
  Sons, 2005.

\bibitem{SM_imperfect}
E.~Basar, U.~Aygolu, E.~Panayirci, and H.~V. Poor, ``Performance of spatial
  modulation in the presence of channel estimation errors,'' \emph{{IEEE}
  Commun. Lett.}, vol.~16, no.~2, pp. 176--179, Feb. 2012.

\bibitem{Zhang_2013}
R.~Zhang, L.~L. Yang, and L.~Hanzo, ``Generalised pre-coding aided spatial
  modulation,'' \emph{IEEE Trans. Wireless Commun.}, vol.~12, no.~11, pp.
  5434--5443, Nov. 2013.

\bibitem{Coon_2019}
\BIBentryALTinterwordspacing
M.-A. Badiu and J.~P. Coon, ``Communication through a large reflecting surface
  with phase errors,'' June 2019. [Online]. Available: \url{arXiv:1906.10751}
\BIBentrySTDinterwordspacing

\end{thebibliography}

\vspace*{-0.53cm}
\begin{IEEEbiography}[{\includegraphics[width=1in,height=1.25in,clip,keepaspectratio]{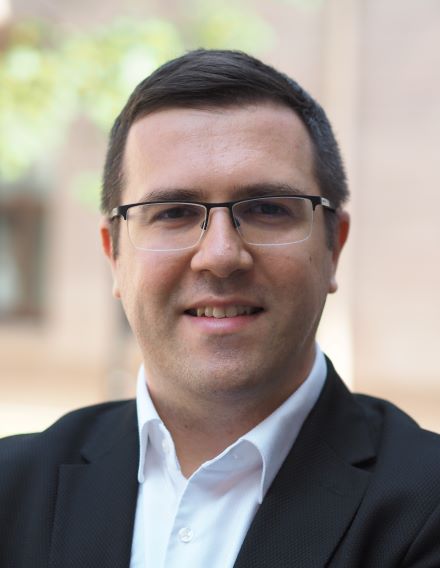}}]{Ertugrul Basar}(S'09-M'13-SM'16) received the B.S. degree (Hons.) from Istanbul University, Turkey, in 2007, and the M.S. and Ph.D. degrees from Istanbul Technical University, Turkey, in 2009 and 2013, respectively. He is currently an Associate Professor with the Department of Electrical and Electronics Engineering, Ko\c{c} University, Istanbul, Turkey and the director of Communications Research and Innovation Laboratory (CoreLab). His primary research interests include MIMO systems, index modulation, intelligent surfaces, waveform design, visible light communications, and signal processing for communications. Dr. Basar currently serves as an Editor of the \textsc{IEEE Transactions on Communications} and \textit{Physical Communication} (Elsevier), and as a Senior Editor of the \textsc{IEEE Communications Letters}.

\end{IEEEbiography}

\end{document}